\documentclass[10pt, twocolumn, twoside]{IEEEtran}

\usepackage{xr}
\usepackage{cite}
\usepackage{enumerate}
\usepackage{graphicx}
\usepackage[cmex10]{amsmath} 
\usepackage{amssymb}
\usepackage{mathtools}
\usepackage{xspace}
\usepackage{subfigure}
\usepackage{multirow}
\usepackage[lined,linesnumbered,ruled,commentsnumbered]{algorithm2e} 
\usepackage{colonequals}
\usepackage[mathscr]{euscript}
\usepackage{amsthm}
\usepackage{mathrsfs}

\usepackage{epsfig}
\usepackage{epstopdf}

\usepackage{tikz}
\usepackage{pgfplots}

\usetikzlibrary{arrows,shapes,backgrounds,plotmarks,positioning}

\pgfplotsset{compat=newest}                         
\pgfplotsset{plot coordinates/math parser=false}
\newlength\figureheight
\newlength\figurewidth

\newtheorem{theorem}{Theorem}[section]

\newtheorem{lemma}[theorem]{Lemma}

\newtheorem{corollary}[theorem]{Corollary}
\newtheorem{proposition}[theorem]{Proposition}

\newtheorem{example}[theorem]{Example}
\newtheorem{remark}[theorem]{Remark}

\newcommand{\argmin}{\arg\!\min}

\newcommand{\op}{\text}
\newcommand{\RZ}[1]{\mathsf{Z}_{#1}}

\newcommand{\RW}[1]{\mathsf{W}_{#1}}
\newcommand{\RCO}{R_{\op{CO}}}
\newcommand{\RACO}{R_{\op{ACO}}}
\newcommand{\RNCO}{R_{\op{NCO}}}
\newcommand{\RRCO}{\mathscr{R}_{\op{CO}}}
\newcommand{\RRACO}{\mathscr{R}_{\op{ACO}}^*}
\newcommand{\RRNCO}{\mathscr{R}_{\op{NCO}}^*}

\newcommand{\XComp}{X_*}
\newcommand{\XCompk}[1]{X_*^{(#1)}}
\newcommand{\TXComp}{\tilde{X}_*}

\newcommand{\alphaComp}{\hat{\alpha}}

\newcommand{\XCompSetk}[1]{\mathcal{X}_*^{(#1)}}
\newcommand{\TXCompSetk}[1]{\tilde{\mathcal{X}}_*^{(#1)}}

\newcommand{\PAR}{\text{PAR}}

\newcommand{\Set}[1]{\{#1\}}

\newcommand{\ASet}[2]{\langle #1 \rangle_{#2}}

\newcommand{\Pat}{\mathcal{P}}
\newcommand{\Qat}[2]{\mathcal{Q}_{#1,#2}}
\newcommand{\X}{\mathcal{X}}
\newcommand{\TX}{\tilde{\mathcal{X}}}

\newcommand{\PiT}{\Pi_{t}}

\newcommand{\U}[2]{\mathcal{U}_{#1,#2}}
\newcommand{\TU}[2]{\tilde{\mathcal{U}}_{#1,#2}}

\newcommand{\Patp}[1]{\mathcal{P}^{(#1)}}
\newcommand{\alphap}[1]{\alpha^{(#1)}}
\newcommand{\alphaU}{\underline{\alpha}}
\newcommand{\alphaNCOp}[1]{\bar{\alpha}^{(#1)}}
\newcommand{\PhiNCO}{\bar{\Phi}}
\newcommand{\phiNCO}{\bar{\phi}}

\newcommand{\rv}{\mathbf{r}} 
\newcommand{\rvNCO}{\bar{\mathbf{r}}}
\newcommand{\rNCO}{\bar{r}}

\newcommand{\Fu}[1]{f_{#1}}
\newcommand{\FuHat}[1]{\hat{f}_{#1}}
\newcommand{\FuU}[1]{g_{#1}}

\newcommand{\Real}{\mathbb{R}}
\newcommand{\Z}{\mathbb{Z}}            

\newcommand{\SFM}{\text{SFM}}

\begin{document}

\title{Part II: A Practical Approach for Successive Omniscience}

\author{Ni~Ding,~\IEEEmembership{Member,~IEEE}, Parastoo~Sadeghi,~\IEEEmembership{Senior Member,~IEEE}, and Thierry~Rakotoarivelo,~\IEEEmembership{Member,~IEEE}

\thanks{Part of the results have been published in \cite{Ding2017ISIT,Ding2018Allerton}.}
\thanks{Ni Ding and Thierry Rakotoarivelo are with Data61 (email: $\{$ni.ding, thierry.rakotoarivelo$\}$@data61.csiro.au).}
\thanks{Parastoo Sadeghi is with the Research School of Electrical, Energy and Materials Engineering (RSEEME), The Australian National University (email: $\{$parastoo.sadeghi$\}$@anu.edu.au). }
}

\markboth{IEEE Transactions}%
{Ding \MakeLowercase{\emph{et al.}}: A Practical Approach for Successive Omniscience}

\maketitle

\begin{abstract}
    In Part I, we studied the communication for omniscience (CO) problem and proposed a parametric ($\PAR$) algorithm to determine the minimum sum-rate at which a set of users indexed by a finite set $V$ attain omniscience.
    The omniscience in CO refers to the status that each user in $V$ recovers the observations of a multiple random source. It is called the global omniscience in this paper in contrast to the study of the successive omniscience (SO), where the local omniscience is attained subsequently in user subsets.
    %
    We apply the $\PAR$ algorithm to search a complimentary subset $\XComp \subsetneq V$ such that
    if the local omniscience in $\XComp$ is reached first, the global omniscience whereafter can still be attained with the minimum sum-rate.
    We further utilize the outputs of the PAR algorithm to outline a multi-stage SO approach that is characterized by $K \leq |V| - 1$ complimentary subsets $\XCompk{k}, \forall k \in \Set{1,\dotsc,K}$ forming a nesting sequence $\XCompk{1} \subsetneq \dotsc \subsetneq \XCompk{K} = V$.
    Starting from stage $k = 1$, the local omniscience in $\XCompk{k}$ is attained at each stage $k$ until the final global omniscience in $\XCompk{K} = V$.
    A $|\XCompk{k}|$-dimensional local omniscience achievable rate vector is also derived for each stage $k$ designating individual users transmitting rates. The sum-rate of this rate vector in the last stage $K$ coincides with the minimized sum-rate for the global omniscience.
\end{abstract}


\section{introduction}

For the users in a finite set $V$ observing distinct terminals of a multiple random source in private, the communication for omniscience (CO) problem in \cite{Csiszar2004} studied how to let users exchange data over broadcast channels to share the knowledge of the entire source.
The \emph{minimum sum-rate problem} in \cite{CourtIT2014,MiloIT2016,Ding2016NetCod,Ding2018IT} aims at the determination of the least sum-rate for users in $V$ to attain omniscience, the state that each user recovers the observation sequence of the entire source, and a corresponding optimal rate vector designating the transmission rates for each user. %
While the CO problem \cite{Csiszar2004} considers the omniscience problem in a one-off manner, the idea of successive omniscience (SO) is proposed in \cite{ChanSuccessive,ChanSuccessiveIT,Ding2015NetCod} revealing that the state of omniscience can be reached in a two-stage manner: let a user subset $X \subsetneq V$ exchange the data first to attain omniscience and the rest of the users overhear the communications; then solve the global omniscience problem in $V$.
By recursively applying the two-stage SO approach, the omniscience in $V$ can be attained in a multi-stage manner.

In \cite{Anoosheh2017SO}, the concept of SO has been applied to the coded cooperative data exchange (CCDE), an application of CO in wireless communications.
A multi-stage SO process was outlined based on a given user subset sequence specifying which group of users to transmit and attain omniscience in each stage.
The problem of determining a local omniscience achievable rate vector for each stage was formulated and solved as a constrained multi-objective optimization problem.
But, it is shown in \cite{ChanSuccessive,ChanSuccessiveIT} that there is a particular group of \emph{complimentary} user subsets so that the local omniscience in any of them can be attained first while the overall communication rates for the global omniscience whereafter still remains minimized.
By knowing that not all user subsets are complimentary, if a non-complimentary subset reaches local omniscience first, e.g., in the solutions to SO in \cite{Anoosheh2017SO}, the users might need to transmit more than the minimum sum-rate to attain the global omniscience finally.
%
Therefore, the \emph{essential problem} in the two-stage SO is not to determine a local omniscience achievable vector, but how to choose a user subset $\XComp \subsetneq V$ that is complimentary
such that the optimality of the global omniscience at the second stage is still maintained.

For a user subset $\XComp \subsetneq V$ to be complimentary, the necessary and sufficient condition was derived in \cite[Theorems 4.2 and 5.2]{ChanSuccessiveIT} for the \emph{asymptotic} and \emph{non-asymptotic models}, where the communication rates are real-valued and integer-valued, respectively.
However, \cite[Theorems 4.2 and 5.2]{ChanSuccessiveIT} are based on the value of the minimum sum-rate for the global omniscience, which is already computationally complex to determine.\footnote{The complexity of the omniscience problem is polynomially growing with the dimension, the number of users $|V|$, of the system \cite{DingITCO2019}. Therefore, attaining omniscience in a user subset is less complex than the global one. This is also a motivation to study the SO problem. }
Meanwhile, the studies on the universal multi-party data exchange problem in \cite{Tyagi2016UniISIT2016,Tyagi2017IT,Tyagi2017ISIT} suggest letting users adaptively increase their transmission rates and running an ideal decoder at the same time to keep searching for the user subset that reaches the omniscience state. The recursive application of this process in \cite[Protocaol~3]{Tyagi2017IT} results in a multi-stage SO.
However, this method requires extra scheduling overheads, e.g., ordering transmission turns based on the information amount (entropy) of individual users' observations and repetitively checking a so-called constant difference property to determine when a user should transmit.
In addition, the ideal decoder needs to be run on line, which also incurs communication overheads between users, e.g., sending ACK/NACK signals.
Thus, the current literature is missing an efficient overall scheduling of the multi-stage SO, before the transmissions actually take place.
More specifically, this scheduling refers to the design of the $K \leq |V| - 1$ stages, for each of which, a complimentary user subset $\XCompk{k}$ that holds the condition in \cite[Theorems 4.2 and 5.2]{ChanSuccessiveIT} is selected and a rate vector $\rv_V^{(k)} = (r_i^{(k)} \colon i \in V)$ is determined with its reduction/projection $\rv_{\XCompk{k}}^{(k)}$ on $\XCompk{k}$ being an achievable local omniscience vector.
In addition, the $\XCompk{K}$ in the last stage must equal $V$ and $\rv_{\XCompk{K}}$ is an optimal rate vector that attains global omniscience with the minimum sum-rate.


\subsection{Contributions}

In this paper, we apply the the $\PAR$ algorithm proposed in Part I \cite{DingITCO2019} to efficiently solve the SO problem for both asymptotic and non-asymptotic source models. 
In each iteration $i$, the $\PAR$ algorithm \cite[Algorithm~2]{DingITCO2019} updates $\Qat{\alpha}{V_i}$, a partition of the users in $V_i$, and a rate vector $\rv_{\alpha,V_i} = (r_{\alpha,i'} \colon i' \in V_i)$ for all values of the minimum sum-rate estimate $\alpha$ for the global omniscience problem.
Here, $V_i$ for $i \in \Set{1,\dotsc,|V|}$ contains the first $i$ users based on an ordering of user indices.
Throughout the $\PAR$ algorithm, the value of the partition $\Qat{\alpha}{V_i}$ is segmented in $\alpha$ and each dimension $r_{\alpha,i}$ of the rate vector $\rv_{\alpha,V_i}$ is piecewise linear in $\alpha$.
At the end of the last iteration $i = |V|$, $\Qat{\alpha}{V}$ and $\rv_{\alpha,V}$ are obtained, where the partition $\Qat{\alpha}{V}$ is segmented in $\alpha$ by $p < |V|$ critical or turning points in $\Set{\alphap{j} \colon j \in \Set{0,\dotsc,p-1}}$.
While Part I \cite{DingITCO2019} applies the value of $\Qat{\alpha}{V}$ and $\rv_{\alpha,V}$ at the first critical point $\alpha = \alphap{1}$ to solve the global omniscience problem,
this paper utilizes $\Qat{\alpha}{V_i}$ and $\rv_{\alpha,V_i}$ at each iteration $i$ to solve the two-stage SO problem and $\Qat{\alpha}{V}$ and $\rv_{\alpha,V}$ in the last iteration to outline a multi-stage SO solution.

We first consider the problem of how to efficiently search a complimentary user subset $\XComp \subsetneq V$ for the two-stage SO. We relax the necessary and sufficient condition in \cite[Theorems 4.2 and 5.2]{ChanSuccessiveIT} to sufficient condition based on a lower bound $\alphaU$ on the minimum sum-rate for the global omniscience. This lower bound can be determined in $O(|V|)$ time.
This sufficient condition is used to prove that, at each iteration $i$ of the $\PAR$ algorithm, any nonsingleton user subset contained in the partition $\Qat{\alphaU}{V_i}$ is complimentary. Here, $\Qat{\alphaU}{V_i}$ is the value of $\Qat{\alpha}{V_i}$ at $\alpha = \alphaU$.
Once the complimentary subset $\XComp$ is chosen as any of the nonsingleton subsets in $\Qat{\alphaU}{V_i}$ at some iteration $i$, a local omniscience achievable rate vector $\rv_{\XComp}$ can be determined simultaneously from $\rv_{\alpha,V_i}$.

We provide two ways for determining a solution to multi-stage SO. The first method is to recursively apply the $\PAR$ algorithm to  choose $\XComp$, let users in $\XComp$ transmit at the rates $\rv_{\XComp}$ and merge them to a super-user after the local omniscience is reached.
Without incurring any transmissions from the users, the second method outlines a multi-stage SO process based on the values of $\Qat{\alpha}{V}$ and $\rv_{\alpha,V}$ at the end of the $\PAR$ algorithm.
For the asymptotic source model, a $p$-stage SO is determined from the critical points as follows. For each stage $k \in \Set{1,\dotsc,p}$, a complimentary $\XCompk{k}$ is extracted as a non-singleton user subset of the partition $\Qat{\alphap{p-k}}{V}$.
By doing so, all $\XCompk{k}$s form a nesting subset sequence $\XCompk{1} \subsetneq \dotsc \subsetneq \XCompk{p}$ such that $\XCompk{p} = V$.
For the complimentary subset $\XCompk{k}$ at stage $k$, a rate vector $\rv_V^{(k)}$ is also determined from $\rv_{\alpha,V}$ with $\rv_{\XCompk{k}}^{(k)}$ designating transmission rates for all users in $\XCompk{k}$ that are sufficient to attain local omniscience in $\XCompk{k}$ and $r_i^{(k)} = 0$ for all $i \in V \setminus \XCompk{k}$ indicating that the rest of users are merely overhearing, i.e., not transmitting.
All $\rv_{V}^{(k)}$s form a nondecreasing sequence $\rv_{V}^{(1)} \leq \dotsc \leq \rv_{V}^{(p)} $ such that $\rv_{V}^{(p)}$ is an optimal rate vector for attaining the global omniscience.
Here, the nesting property of $\XCompk{k}$ and the monotonicity of $\rv_{V}^{(k)}$ guarantee the achievability of this multi-stage SO in practice: start from stage $k = 1$; at each stage $k \in \Set{2,\dotsc,p}$, a larger user subset $\XCompk{k}$ attains local omniscience by the nonnegative transmission rates $\rv_V^{(k)} - \rv_V^{(k-1)}$.
The number of users reaching omniscience $|\XCompk{k}|$ increases with $k$ until the global omniscience in $V$ is attained in the last stage $k = p$.

Similarly, for the non-asymptotic model, a $K$-stage SO with $K \leq p$ can be determined by the $\PAR$ algorithm: $\XCompk{k}$ is extracted as a non-singleton user subset of the partition $\Qat{\alpha}{V}$ at the integer-valued $\alpha \in  \Set{ \lceil \alphap{p-k} \rceil \colon k \in \Set{1,\dotsc,p}}$. The transmission rate vector $\rv_V^{(k)}$ for each $k$ can be determined by one more call of the $\PAR$ algorithm.
The study in this paper shows that the SO problem in both asymptotic and non-asymptotic models can be solved by the $\PAR$ algorithm in $O(|V| \cdot \SFM(|V|))$ time. Here, $\SFM(|V|)$ denotes the complexity of solving a submodular function minimization (SFM) problem and is a polynomial function of $|V|$.

\subsection{Organization}

The rest of paper is organized as follows. The system model and the SO problem are described in Section~\ref{sec:systemSO}.
Section~\ref{sec:XComp} derives the sufficient condition for a user subset to be complimentary and Section~\ref{sec:1stSO} shows how to implement the two-stage SO in both asymptotic and non-asymptotic models. In Section~\ref{sec:MstSO}, we show how to extract the multi-stage SO procedures from $\Qat{\alpha}{V}$ and $\rv_{\alpha,V}$ for asymptotic and non-asymptotic models.

\section{System Model}
\label{sec:systemSO}

Let the \emph{ground set} $V$ with $|V|>1$ contain all users in the system. The multiple random source is denoted by $\RZ{V}=(\RZ{i}:i\in V)$ with each $\RZ{i}$ being a discrete random variable. User $i$ privately observes an $n$-sequence $\RZ{i}^n$ of $\RZ{i}$. The users are allowed to exchange their data to help each other recover the observation of the source.
For a user subset $X \subseteq V$, the state that each user in $X$ recovers the observation sequence $\RZ{X}^n$ of $\RZ{X}$ is called the \emph{local omniscience} in $X$. In the case when $X = V$, we say that the \emph{global omniscience} is attained.

For the local omniscience in the user subset $X \subseteq V$, we briefly summarize the results on the CO problem in \cite[Section~\ref{subsec:MinSumRate}]{DingITCO2019}.
For a vector $\rv_X=(r_i \colon i \in X)$ with each $r_i$ denoting the rate at which user $i$ broadcasts/transmits, $\rv_X$ is called a \emph{local omniscience achievable rate vector} if all the users in $X$ are able to recover $\RZ{X}^n$ after transmitting at the rates $\rv_X$.
The corresponding \emph{local omniscience achievable rate region} is \cite{Csiszar2004,SW1973,Cover1975}:
$$ \RRCO(X)=\Set{ \rv_X \in \Real^{|X|} \colon r(C) \geq H(C|V\setminus C),\forall C \subsetneq X },$$
where $r(C)=\sum_{i\in C} r_i$ is the \emph{sum-rate} in the subset $C$ of $X$.
In an \emph{asymptotic model}, we consider the asymptotic limits as the \emph{block length} $n$ goes to infinity so that the communication rates could be real or fractional; In a non-asymptotic model, the block length $n$ is restricted to be finite and the communication rates are required to be integral.

The minimum sum-rates for attaining the local omniscience in $X$ in the asymptotic and non-asymptotic models are $\RACO(X)=\min\Set{r(X) \colon \rv_X\in \RRCO(X)}$ and $\RNCO(X)=\min\Set{r(X) \colon \rv_X\in \RRCO(X) \cap \Z^{|X|} }$, respectively. The corresponding optimal rate vector sets are $\RRACO(X)=\Set{\rv_X\in \RRCO(X) \colon r(X)=\RACO(X)} $ and $\RRNCO(X)=\Set{\rv_X \in \RRCO(X) \cap \Z^{|X|} \colon r(X)=\RNCO(X)}$ for the asymptotic and non-asymptotic models, respectively.
Let $\Pi(X)$ be the set of all partitions of $X$. It is shown in  \cite[Example 4]{Csiszar2004} \cite{Chan2008tight} \cite[Corollary 6]{Ding2018IT} that
    \begin{equation} \label{eq:MinSumRateSO}
            \RACO(X) = \max_{\Pat \in \Pi(X) \colon |\Pat| > 1} \sum_{C \in \Pat} \frac{H(X) - H(C)}{|\Pat|-1}
    \end{equation}
and $\RNCO(X) = \lceil \RACO(X) \rceil$.

\subsection{Successive Omniscience}

The concept of SO is outlined in \cite{ChanSuccessiveIT,Ding2015NetCod}. It is shown that, instead of the one-off approach in \cite{Roua2010,Court2010,Court2010M,Ding2018IT,CourtIT2014,MiloIT2016}, the communications between the users in $V$ can be organized in a way such that global omniscience in $V$ is attained in a two-stage manner.
First, let the users in a subset $X$ broadcast to attain the local omniscience and the remaining users $i \in V \setminus X$ overhear these transmissions; Then, solve the global omniscience problem in $V$.
This two-stage approach can be implemented in a way without losing the optimality of the global omniscience: it is shown in \cite{ChanSuccessiveIT}  that there is a particular group of nonsingleton subsets $\XComp$ such that the local omniscience in $\XComp$ can be attained first so that the overall sum-rate for attaining the global omniscience in $V$ still remains minimized. We call $\XComp$ a \emph{complimentary subset} \cite{ChanSuccessiveIT}.

Thus, the problem of SO boils down to searching the complimentary user subset $\XComp$ and determining the local omniscience achievable rate vectors $\rv_{\XComp} \in \RRCO(\XComp)$
and $\rv_{\XComp} \in \RRCO(\XComp) \cap \Z^{|\XComp|}$ for the asymptotic and non-asymptotic models, respectively.
In a multi-stage SO, the problem is to determine the complimentary user subset $\XCompk{k}$ and the local omniscience achievable rate vectors $\rv_{\XCompk{k}}^{(k)} \in \RRCO(\XCompk{k})$
and $\rv_{\XCompk{k}}^{(k)} \in \RRCO(\XCompk{k}) \cap \Z^{|\XCompk{k}|}$ for the asymptotic and non-asymptotic models, respectively, for each stage $k \in \Set{1,\dotsc,K-1}$ and
ensure $\XCompk{K} = V$ and $\rv_{V}^{(K)} \in \RRACO(V)$ and $\rv_{V}^{(K)} \in \RRNCO(V)$ for the asymptotic and non-asymptotic models, respectively, in last stage $K$.
This paper shows how to solve the two-stage and multi-stage SO by the $\PAR$ algorithm proposed in Part I \cite{DingITCO2019}. Before presenting the solutions in Sections~\ref{sec:XComp} to \ref{sec:MstSO}, we briefly review the exiting results on the global omniscience problem in \cite{ChanMMI,Ding2016NetCod,Ding2018IT} and the $\PAR$ algorithm below.

      \begin{algorithm} [t]
	       \label{algo:ParAlgoAux}
	       \small
	       \SetAlgoLined
	       \SetKwInOut{Input}{input}\SetKwInOut{Output}{output}
	       \SetKwFor{For}{for}{do}{endfor}
            \SetKwRepeat{Repeat}{repeat}{until}
            \SetKwIF{If}{ElseIf}{Else}{if}{then}{else if}{else}{endif}
	       \BlankLine
           \Input{$H$, $V$ and $\Phi$}
	       \Output{segmented variables $\rv_{\alpha,V} \in B(\FuHat{\alpha})$ and $\Qat{\alpha}{V} = \bigwedge \argmin_{\Pat\in \Pi(V)} \Fu{\alpha}[\Pat]$ for all $\alpha$}
	       \BlankLine
            $\rv_{\alpha,V} \coloneqq (\alpha - H(V)) \chi_V$ for all $\alpha$\;
            $ r_{\alpha,\phi_1} \coloneqq \Fu{\alpha}(\Set{\phi_1})$ and $\Qat{\alpha}{V_1} \coloneqq \Set{\Set{\phi_1}}$ for all $\alpha$\;
            \For{$i=2$ \emph{\KwTo} $|V|$}{
                $\Qat{\alpha}{V_i} \coloneqq \Qat{\alpha}{V_{i-1}} \sqcup \Set{\Set{\phi_i}}$ for all $\alpha$\;
                Obtain the minimal minimizer $\U{\alpha}{V_i}$ of $\min\Set{ \FuU{\alpha}(\TX) \colon \Set{\phi_i} \in \X \subseteq \Qat{\alpha}{V_i}}$ for all $\alpha$\;
                For all $\alpha$, update $\rv_V$ and $\Qat{\alpha}{V_i}$ by \label{step:UpdatesParAux}
                \begin{equation}
                    \begin{aligned}
                        \rv_{\alpha,V_i} &\coloneqq \rv_{\alpha,V_i} + \FuU{\alpha}(\TU{\alpha}{V_i}) \chi_{\phi_i}; \\
                        \Qat{\alpha}{V_i} &\coloneqq (\Qat{\alpha}{V_i} \setminus \U{\alpha}{V_i}) \sqcup \Set{ \TU{\alpha}{V_i} };
                    \end{aligned} \nonumber
                \end{equation}
                for all $\alpha$;
            }
            \Return $\rv_V$ and $\Qat{\alpha}{V}$ for all $\alpha$\;
	   \caption{Parametric (PAR) Algorithm \cite[Algorithm~\ref{algo:ParAlgo}]{DingITCO2019}}
	   \end{algorithm}

\subsection{The $\PAR$ Algorithm}
\label{subsec:PARReview}

For a given minimum sum-rate estimate $\alpha \in [0, H(V)]$, the \emph{residual entropy function} is $\Fu{\alpha}(X) = \alpha - H(V) +H(X), \forall X \subseteq V$. 
The solution to the global omniscience problem in $V$ is closely related to the \emph{Dilworth truncation} of $\Fu{\alpha}$: $\FuHat{\alpha}(V) =  \min_{\Pat \in \Pi(V)} \Fu{\alpha}[\Pat]$, where $\Fu{\alpha}[\Pat] = \sum_{C \in \Pat} \Fu{\alpha}(C)$.
In Part I \cite[Algorithm~\ref{algo:ParAlgo}]{DingITCO2019}, we proposed a parametric ($\PAR$) algorithm (see Algorithm~\ref{algo:ParAlgoAux}), where the iteration are run in the order of an arbitrary \emph{linear ordering/permutation} $\Phi = (\phi_1,\dotsc,\phi_{|V|})$ of user indices.

It is shown in \cite[Proposition~\ref{prop:preamble}]{DingITCO2019} that, for $V_i = \Set{\phi_1,\dotsc,\phi_i}$ that contains the first $i$ users in $\Phi$, the finest minimizer $\Qat{\alpha}{V_i} = \bigwedge \argmin_{\Pat \in \Pi(V_i)} \Fu{\alpha}[\Pat]$ for all $\alpha$ segmented as\footnote{As in Part I \cite{DingITCO2019}, `for all $\alpha$' means `for all $\alpha \in [0, H(V)]$' in this paper.}
\begin{equation}
    \Qat{\alpha}{V_i} = \begin{cases}
                        \Patp{p} & \alpha \in [0,\alphap{p}],\\
                        \Patp{p-1} & \alpha \in (\alphap{p},\alphap{p-1}],\\
                        &\vdots \\
                        \Patp{0} & \alpha \in (\alphap{1},\alphap{0}]
                     \end{cases}   \nonumber
\end{equation}
is obtained at end of each iteration $i$ of the $\PAR$ algorithm, where $0 \leq \alphap{p} < \dotsc < \alphap{0} = H(V)$ and $\Set{\Set{m} \colon m \in V_i} = \Patp{p} \prec \dotsc \prec \Patp{1} \prec \Patp{0} = \Set{V_i}$. Here, $\Pat \prec \Pat'$ denotes $\Pat$ is \emph{strictly finer than} $\Pat'$.
The critical points $\alphap{j}$s and partitions $\Patp{j}$s constitute the \emph{principal sequence of partitions (PSP)} of $V_i$.
The $\PAR$ algorithm finally outputs the PSP of $V$ that segments $\Qat{\alpha}{V} = \bigwedge \argmin_{\Pat \in \Pi(V)} \Fu{\alpha}[\Pat]$, where the first critical/turning point refers to $\alphap{1}$ and equals the minimum sum-rate $\RACO(V)$ for attaining the global omniscience. The corresponding value of $\Qat{\alphap{1}}{V} = \Patp{1}$ is the finest maximizer of \eqref{eq:MinSumRateSO} for $X = V$ and is called the \emph{fundamental partition}.

The $\PAR$ algorithm also returns a rate vector $\rv_{\alpha,V} = (r_{\alpha,i} \colon i \in V) $ that is piecewise linear in $\alpha$ such that $\rv_{\alpha,V} \in B(\FuHat{\alpha})$ for all $\alpha$, where $B(\FuHat{\alpha}) = \Set{\rv_V \in P(\FuHat{\alpha}) \colon r(V) = \FuHat{\alpha}(V)}$ and $P(\FuHat{\alpha}) = \Set{\rv_V \in \Real \colon r(X) \leq \FuHat{\alpha}(X), X \subseteq V}$ is the \emph{base polyhedron} and \emph{polyhedron} of $\FuHat{\alpha}$, respectively.
Due to the equivalence, $B(\FuHat{\alpha}) = \Set{\rv_{V} \in \RRCO(V) \colon r(V) = \alpha}$ for all $\alpha \geq \RACO(V)$ \cite[Section~III-B and Theorem~4]{Ding2018IT}, $\rv_{\RACO(V),V} \in \RRACO(V)$ and $\rv_{\RNCO(V),V} \in \RRNCO(V)$ are the optimal rate vectors for asymptotic and non-asymptotic models, respectively.
The $\PAR$ algorithm completes in $O(|V| \cdot \SFM(|V|))$ time, where $\SFM(|V|)$ denotes the complexity of minimizing a submodular function $\min\Set{f(X) \colon X \subseteq V}$.\footnote{Function $f \colon 2^V \mapsto \Real$ is \emph{submodular} if $f(X) + f(Y) \geq f(X \cap Y) + f(X \cup Y)$ for all $X,Y \subseteq V$.}

\section{Complimentary Subset}
\label{sec:XComp}

In this section, we show that the existence and non-existence of a complimentary subset can be determined by a lower bound on the minimum sum-rate, $\RACO(V)$ and $\RNCO(V)$ for asymptotic and non-asymptotic models, respectively. When this lower bound is applied to $\Qat{\alpha}{V_i}$ and $\rv_{\alpha,V_i}$ obtained at the end of each iteration $i$ of the $\PAR$ algorithm, a complimentary subset $\XComp$ and a local omniscience achievable rate vector $\rv_{\XComp}$ are both determined for SO.

In \cite{ChanSuccessive,ChanSuccessiveIT}, the authors derived the necessary and sufficient condition for a user subset to be complimentary for both asymptotic and non-asymptotic models.

\begin{theorem}[necessary and sufficient condition {\cite[Theorems 4.2 and 5.2]{ChanSuccessiveIT}}]\label{theo:SOIff}
    In an asymptotic model, a user subset $\XComp \subsetneq V$ such that $|\XComp| > 1$ is complimentary if and only if $H(V) - H(\XComp) + \RACO(\XComp) \leq \RACO(V)$; In a non-asymptotic model, a user subset $\XComp \subsetneq V$ such that $|\XComp| > 1$ is complimentary if and only if $H(V) - H(\XComp) + \RNCO(\XComp) \leq \RNCO(V)$.\footnote{ In {\cite[Theorems 4.2 and 5.2]{ChanSuccessiveIT}}, the necessary and sufficient condition for $X$ to be complimentary is $I(X) \geq I(V)$ and $\lfloor I(X) \rfloor \geq \lfloor I(V) \rfloor$ for the asymptotic and non-asymptotic models, respectively, which are converted to the ones in Theorem~\ref{theo:SOIff} via the dual relationships: $\RACO(V) = H(V)- I(V) $ and $\RNCO(V) = H(V)- \lfloor I(V) \rfloor$ \cite{ChanMMI,Court2011}. Here, $I(V)$ is the amount of information shared by users in $V$. See \cite[Section~\ref{sec:Relation}]{DingITCO2019}.
    Since $V$ is always a complimentary subset, we restrict our attention to proper subsets of $V$.} \hfill\IEEEQED
\end{theorem}

In Theorem~\ref{theo:SOIff}, $H(V) - H(\XComp)$ is the amount of information that is missing in user subset $\XComp$, the omniscience of which only relies on the transmissions from the users in $V \setminus \XComp$. If we let the users in $\XComp$ attain local omniscience with the minimum sum-rate $\RACO(\XComp)$, the users in $V \setminus \XComp$ are required to transmit at least at the rate $H(V) - H(\XComp)$ for attaining the global omniscience. Then, the total number of transmissions is no less than $H(V) - H(\XComp) + \RACO(\XComp)$. On the contrary, if $H(V) - H(\XComp) + \RACO(\XComp)>\RACO(V)$, the global omniscience is not achievable by the minimum sum-rate $\RACO(V)$ if we allow the users in $\XComp$ to attain the local omniscience first. The condition $H(V) - H(\XComp) + \RNCO(\XComp) \leq \RNCO(V)$ for the non-asymptotic model in Theorem~\ref{theo:SOIff} can be interpreted in the same way.

The necessary and sufficient condition in Theorem~\ref{theo:SOIff} assume that the value of $\RACO(V)$ or $\RNCO(V)$ is known in advance, which requires solving the global omniscience problem first. It is not practical to directly apply Theorem~\ref{theo:SOIff} in that one of the motivations of SO is to avoid dealing with the large-scale omniscience problem: the complexity $O(|V| \cdot \SFM(|V|))$ grows with the number of users so that the local omniscience problem can be solved faster than the global one.
In addition, except for a brute-force search over the power set $2^V$, Theorem~\ref{theo:SOIff} does not provide any practical method for efficiently searching for the complimentary subset $\XComp$.

\subsection{A Sufficient Condition}

We rewrite Theorem~\ref{theo:SOIff} in terms of the Dilworth truncation $\FuHat{\alpha}$ below and relax it to a sufficient condition that only requires a lower bound on the minimum sum-rate $\RACO(V)$ or $\RNCO(V)$. The proof of Corollary~\ref{coro:SOEqv} is in Appendix~\ref{app:coro:SOEqv}.

\begin{corollary} \label{coro:SOEqv}
    In an asymptotic model, a user subset $\XComp \subsetneq V$ such that $|\XComp| > 1$ is complimentary if and only if $\Fu{\RACO(V)}(\XComp) = \FuHat{\RACO(V)}(\XComp)$; In a non-asymptotic model, a user subset $\XComp \subsetneq V$ such that $|\XComp| > 1$ is complimentary if and only if $\Fu{\RNCO(V)}(\XComp) = \FuHat{\RNCO(V)}(\XComp)$. \hfill \IEEEQED
\end{corollary}

\begin{lemma}[sufficient condition]\label{lemma:SOSuf}
    In an asymptotic model, a user subset $\XComp \subsetneq V$ such that $|\XComp| > 1$ is complimentary if $\Fu{\alphaU}(\XComp) = \FuHat{\alphaU}(\XComp)$ for $\alphaU \leq \RACO(V)$; In a non-asymptotic model, a user subset $\XComp \subsetneq V$ such that $|\XComp| > 1$ is complimentary if $\Fu{\alphaU}(\XComp) = \FuHat{\alphaU}(\XComp)$ for an integer-valued $\alphaU \leq \RNCO(V)$.
\end{lemma}
\begin{IEEEproof}
    For any $\alphaU, \alphaU' \leq \RACO(V)$ such that $\alphaU < \alphaU'$, if $\Fu{\alphaU}(\XComp) = \FuHat{\alphaU}(\XComp)$, then
    \begin{equation} \label{eq:lemma:SOSuf:aux}
        \begin{aligned}
            & \Fu{\alphaU'}[\Pat] - \Fu{\alphaU'}(\XComp) \\
            & \qquad = H[\Pat] - H(\XComp) - (|\Pat| - 1) \Big( H(V) - \alphaU' \Big) \\
            & \qquad > H[\Pat] - H(\XComp) - (|\Pat| - 1) \big( H(V) - \alphaU \big) \\
            & \qquad = \Fu{\alphaU}[\Pat] - \Fu{\alphaU}(\XComp) \geq 0
        \end{aligned}
    \end{equation}
    for all $\Pat \in \Pi(\XComp)$ such that $|\Pat| > 1$, where $H[\Pat] = \sum_{C \in \Pat} H(C)$. So, $\Fu{\RACO(V)}[\Pat] - \Fu{\RACO(V)}(\XComp) \geq \Fu{\alphaU}[\Pat] - \Fu{\alphaU}(\XComp) \geq 0, \forall\Pat \in \Pi(\XComp) \colon |\Pat| > 1$. The condition $\Fu{\RACO(V)}(\XComp) = \FuHat{\RACO(V)}(\XComp)$ in Corollary~\ref{coro:SOEqv} holds. Therefore, $\XComp$ is complimentary in the asymptotic model.
    In the same way, one can prove the sufficient condition $\Fu{\alphaU}(\XComp) = \FuHat{\alphaU}(\XComp)$ for the non-asymptotic model.
\end{IEEEproof}

It is not difficult to obtain a lower bound on the minimum sum-rate for the global omniscience problem. According to \cite[Proposition~14]{Ding2018IT}, a possible value for $\alphaU$ in Corollary~\ref{coro:SOEqv} can be determined as the value of objective function in \eqref{eq:MinSumRateSO} over only the singleton partition and the bi-partitions in $\PiT(V) = \big\{ \Set{\Set{m} \colon m \in V}, \Set{\Set{i},V \setminus \Set{i}} \colon i \in V \big\}$:
\begin{subequations}\label{eq:LB}
    \begin{align}
        & \alphaU = \max_{\Pat \in \PiT(V)} \sum_{C \in \Pat} \frac{H((V) - H(C))}{|\Pat| - 1}  \leq \RACO(V),  \label{eq:LBACO}\\
        & \alphaU = \Big\lceil \max_{\Pat \in \PiT(V)} \sum_{C \in \Pat} \frac{H((V) - H(C))}{|\Pat| - 1} \Big\rceil \leq \RNCO(V)  \label{eq:LBNCO}
    \end{align}
\end{subequations}
for the asymptotic and non-asymptotic models, respectively. The lower bound in \eqref{eq:LB} can be determined by $O(|V|)$ calls of the entropy function.

\begin{example} \label{ex:SOSuf}
    For the $5$-user system in \cite[Example~\ref{ex:main}]{DingITCO2019} where
    \begin{equation}
        \begin{aligned}
            \RZ{1} & = (\RW{b},\RW{c},\RW{d},\RW{h},\RW{i}),   \\
            \RZ{2} & = (\RW{e},\RW{f},\RW{h},\RW{i}),   \\
            \RZ{3} & = (\RW{b},\RW{c},\RW{e},\RW{j}), \\
            \RZ{4} & = (\RW{a},\RW{b},\RW{c},\RW{d},\RW{f},\RW{g},\RW{i},\RW{j}),  \\
            \RZ{5} & = (\RW{a},\RW{b},\RW{c},\RW{f},\RW{i},\RW{j}),
        \end{aligned}  \nonumber
    \end{equation}
    with each $\RW{m}$ being an independent and uniformly distributed random bit, it can be shown that
     \begin{equation}
        \begin{aligned}
            &\Set{\XComp \subsetneq V \colon |\XComp| > 1, H(V) - H(\XComp) + \RACO(\XComp) \leq \RACO(V)} \\
            &\qquad = \Set{\XComp \subsetneq V \colon |\XComp|>1, \Fu{\RACO(V)}(\XComp) = \FuHat{\RACO(V)}(\XComp) } \\
            &\qquad = \big\{ \Set{4,5},\Set{1,4},\Set{1,4,5},\Set{1,2,3,4} \big\},
        \end{aligned} \nonumber
     \end{equation}
     are all complimentary subsets for the asymptotic model and
     \begin{equation}
        \begin{aligned}
            &\Set{\XComp \subsetneq V \colon |\XComp| > 1, H(V) - H(\XComp) + \RNCO(\XComp) \leq \RNCO(V)} \\
            & \qquad = \Set{ \XComp \subsetneq V \colon |\XComp|>1, \Fu{\RNCO(V)}(\XComp) = \FuHat{\RNCO(V)}(\XComp) } \\
            & \qquad = \big\{ \Set{1,4},\Set{1,5},\Set{3,4},\Set{3,5},\Set{4,5},\Set{1,2,4},\\
            & \qquad\qquad \Set{1,2,5},\Set{1,3,4},\Set{1,3,5},\Set{1,4,5},\Set{2,3,4},\\
            & \qquad\qquad \Set{2,3,5},\Set{3,4,5},\Set{1,2,3,4},\Set{1,2,3,5},\\
            & \qquad\qquad \Set{1,2,4,5},\Set{1,3,4,5},\Set{2,3,4,5} \big\},
        \end{aligned} \nonumber
     \end{equation}
     are all complimentary subsets for the non-asymptotic model.

    Instead, a lower bound on $\RACO(V)$ can be obtained as $\alphaU = \sum_{i \in V} \frac{H((V) - H(\Set{i}))}{|\Pat| - 1} = 5.75  < \RACO(V)$, where we apply Lemma~\ref{lemma:SOSuf} to find a complimentary subset
    $$ \Set{ \XComp \subsetneq V \colon |\XComp|>1, \Fu{\alphaU}(\XComp) = \FuHat{\alphaU}(\XComp) } = \Set{\Set{4,5}}  $$
    for the asymptotic model.
    Consider a tighter lower bound $\alphaU = \max_{\Pat \in \PiT(V)} \sum_{C \in \Pat} \frac{H((V) - H(C))}{|\Pat| - 1} = 6 $ on both $\RACO(V)$ and $\RNCO(V)$. We have
    \begin{multline} 
      \Set{ \XComp \subsetneq V \colon |\XComp|>1, \Fu{\alphaU}(\XComp) = \FuHat{\alphaU}(\XComp) } \\ = \Set{\Set{1,4},\Set{4,5},\Set{1,4,5}} \nonumber
    \end{multline}
    being the complimentary subsets for both asymptotic and non-asymptotic models.
\end{example}

\subsection{Existence of A Complimentary Subset}

There are two observations in Example~\ref{ex:SOSuf} that need to be explored further. First, the number of complimentary subsets decreases with the value of lower bound $\alphaU$. This is proved by the corollary below.

\begin{corollary} \label{coro:SOSufLB}
    For any two lower bounds $\alphaU$ and $\alphaU'$ on the minimum sum-rate $\RACO(V)$ for the asymptotic model, or on $\RNCO(V)$ for the non-asymptotic model, such that $\alphaU < \alphaU'$,
    \begin{multline}
        \Set{ \XComp \subsetneq V \colon |\XComp|>1, \Fu{\alphaU}(\XComp) = \FuHat{\alphaU}(\XComp) } \\
            \subseteq \Set{ \XComp \subsetneq V \colon |\XComp|>1, \Fu{\alphaU'}(\XComp) = \FuHat{\alphaU'}(\XComp) }. \nonumber
    \end{multline}
\end{corollary}
\begin{IEEEproof}
    As shown in the proof of Lemma~\ref{lemma:SOSuf}, for any $\XComp \subsetneq V$ such that $|\XComp| > 1$, if $\Fu{\alphaU}(\XComp) = \FuHat{\alphaU}(\XComp)$, then $\Fu{\alphaU}[\Pat] - \Fu{\alphaU}(\XComp) \geq 0, \forall \Pat \in \Pi(\XComp) \colon |\Pat| > 1$ and inequality \eqref{eq:lemma:SOSuf:aux} holds. We necessarily have $\Fu{\alphaU'}(\XComp) = \FuHat{\alphaU'}(\XComp)$ so that $\XComp \in \Set{ \XComp \subsetneq V \colon |\XComp|>1, \Fu{\alphaU'}(\XComp) = \FuHat{\alphaU'}(\XComp)}$. Corollary holds.
\end{IEEEproof}

Corollary~\ref{coro:SOSufLB} means that all complimentary subsets for the asymptotic model are also complimentary in the non-asymptotic model (See Example~\ref{ex:SOSuf}) since $\RACO(V) \leq \RNCO(V)$. The second observation in Example~\ref{ex:SOSuf} is also based on Corollary~\ref{coro:SOSufLB}: Lemma~\ref{lemma:SOSuf} just allows us to search a part of, rather than all, the complimentary subsets. As the number of complimentary subsets searched by Lemma~\ref{lemma:SOSuf} shrinks to zero when $\alphaU$ decreases, we should choose a lower bound $\alphaU$ large enough to capture at least one of the complimentary subsets, if there exists one.
For this purpose, the following lemma states that, for the lower bounds $\alphaU = \sum_{i \in V} \frac{H(V) - H(\Set{i})}{|V| - 1}$ and $\alphaU = \big\lceil \sum_{i \in V} \frac{H(V) - H(\Set{i})}{|V| - 1} \big\rceil$ determined by the singleton partition $\Set{\Set{i} \colon i \in V}$ for the asymptotic and non-asymptotic models, respectively, the sufficient condition in Lemma~\ref{lemma:SOSuf} is enough to prove the nonexistence of the complimentary subset.

\begin{lemma} \label{lemma:SONXComp}
    There does not exist any complimentary subset in the asymptotic model if no $\XComp \subsetneq V$ such that $|\XComp| > 1$ satisfies $\Fu{\alphaU}(\XComp) = \FuHat{\alphaU}(\XComp)$ for $\alphaU = \sum_{i \in V} \frac{H(V) - H(\Set{i})}{|V| - 1}$; There does not exist any complimentary subset in the non-asymptotic model if no $\XComp \subsetneq V$ such that $|\XComp| > 1$ satisfies $\Fu{\alphaU}(\XComp) = \FuHat{\alphaU}(\XComp)$ for $\alphaU = \big\lceil \sum_{i \in V} \frac{H(V) - H(\Set{i})}{|V| - 1} \big\rceil$. \hfill\IEEEQED
\end{lemma}

The proof of Lemma~\ref{lemma:SONXComp} is in Appendix~\ref{app:lemma:SONXComp}.
An example to demonstrate Lemma~\ref{lemma:SONXComp} for the asymptotic model is the independent source model with the terminals $\RZ{i}$ being independent of each other, where we have only two trivial partitions $\Patp{1} = \Set{\Set{i} \colon i \in V}$ and $\Patp{0} = \Set{V}$ in the PSP and $\alphap{1} = \RACO(V) = H(V)$. The lower bound in Lemma~\ref{lemma:SONXComp} is $\alphaU = \sum_{i \in V} \frac{H(V) - H(\Set{i})}{|V| - 1} = \alphap{1}$ and the partition $\Patp{1} = \Set{\Set{i} \colon i \in V}$ indicates that no $\XComp \subsetneq V$ such that $|\XComp| > 1$ satisfies $\Fu{\alphaU}(\XComp) = \FuHat{\alphaU}(\XComp)$. Therefore, there does not exist any complimentary subset for the asymptotic model in the independent source.
Lemmas~\ref{lemma:SONXComp} and \ref{lemma:SOSuf} suggest an efficient method for searching the complimentary subset by the lower bound $\alphaU$, which is applied to the two-stage SO in Section~\ref{sec:1stSO} and the multi-stage SO in Section~\ref{sec:MstSO}.

\section{Two-stage Successive Omniscience}
\label{sec:1stSO}

The original SO problem in \cite{ChanSuccessive,ChanSuccessiveIT} is outlined in two steps: select a complimentary subset $\XComp$ and attain local omniscience in it; fuse all $i \in \XComp$ into one super-user and consider the omniscience problem in the new system containing the fused super-user $\TXComp$ and the rest of the users in $i \in V \setminus \XComp$.
This two-stage SO approach can be applied recursively so that a sequence of local omniscience leads to the global one and hence the name `successive'.
It is clear that the two-stage SO problem is solved if a complimentary $\XComp$ is searched and the local omniscience problem in it is solved.
While the sufficient condition in Lemma~\ref{lemma:SOSuf} only determine a $\XComp$, we show in this section that, when it is applied to the segmented $\Qat{\alpha}{V_i}$ and $\rv_{\alpha,V_i}$ in the $\PAR$ algorithm, not only a $\XComp$, but also a local omniscience achievable rate vector $\rv_{\XComp}$ can be determined at the same time for both asymptotic and non-asymptotic models.

The following lemma states that the existence of the complimentary subset and at least one of the complimentary subsets, if there exist one, can be determined by applying the lower bound $\alphaU$ in Lemma~\ref{lemma:SONXComp} to the $\PAR$ algorithm. A local omniscience achievable rate vector is also obtained at the same time.
The proof is in Appendix~\ref{app:lemma:SOPAR}.

%

\begin{lemma} \label{lemma:SOPAR}
    Let $\Qat{\alpha}{V_i}$ and $\rv_{\alpha,V_i}$ be the segmented partition and rate vector, respectively, obtained at the end of any iteration $i \in \Set{2,\dotsc,|V|}$ of the PAR algorithm.
    \begin{enumerate}[(a)]
        \item For the asymptotic model, let $\alphaU = \sum_{i \in V} \frac{H(V) - H(\Set{i})}{|V| - 1}$. Any nonsingleton $C \in \Qat{\alphaU}{V_i}$ is a complimentary subset and $\rv_{\alphaComp,C}$ for $\alphaComp = \min \Set{ \alpha \in \Real \colon \Fu{\alpha} (C) = \FuHat{\alpha} (C)}$ is an optimal rate vector that attains local omniscience in $C$ with the minimum sum-rate $\RACO(C)$.
        \item For the non-asymptotic model, let $\alphaU = \big\lceil \sum_{i \in V} \frac{H(V) - H(\Set{i})}{|V| - 1} \big\rceil$. Any nonsingleton $C \in \Qat{\alphaU}{V_i}$ is a complimentary subset and $\rv_{\alphaComp,C}$ for $\alphaComp = \min \Set{ \alpha \in \Z \colon \Fu{\alpha} (C) = \FuHat{\alpha} (C)}$ is an optimal rate vector that attains local omniscience in $C$ with the minimum sum-rate $\RNCO(C)$.
    \end{enumerate}
    For both asymptotic and non-asymptotic models, if all subsets in $\Qat{\alphaU}{V_i}$ remain singleton $\Qat{\alphaU}{V_i} = \Set{\Set{m} \colon m \in V_i}$ until the $|V|$th iteration, there does not exist a complimentary subset. \hfill\IEEEQED
\end{lemma}

Lemma~\ref{lemma:SOPAR} is implemented by Algorithm~\ref{algo:1stSO}. When there does not exist a complimentary subset, Algorithm~\ref{algo:1stSO} outputs the results $\Qat{\alpha}{V}$ and $\rv_{\alpha,V}$ determining the minimum sum-rate and an optimal rate vector for the global omniscience problem in $V$.
That is, if there exists a complimentary subset, Algorithm~\ref{algo:1stSO} determines one such subset $\XComp$ and local omniscience rate vector $\rv_{\XComp}$ for the two-stage SO in $O(|V| \cdot \SFM(|V|))$ time; otherwise, the minimum sum-rate problem in $V$ is solved in $O(|V| \cdot \SFM(|V|))$ time.

        \begin{algorithm} [t]
	       \label{algo:1stSO}
	       \small
	       \SetAlgoLined
	       \SetKwInOut{Input}{input}\SetKwInOut{Output}{output}
	       \SetKwFor{For}{for}{do}{endfor}
            \SetKwRepeat{Repeat}{repeat}{until}
            \SetKwIF{If}{ElseIf}{Else}{if}{then}{else if}{else}{endif}
	       \BlankLine
           \Input{$H$, $V$ and (an arbitrarily chosen linear ordering) $\Phi$.}
	       \Output{a complimentary subset $C$ and an optimal rate vector $\rv_{\alphaComp,C}$ for attaining local omniscience in $C$; Or, an empty set indicating no complimentary subset and $\Qat{\alpha}{V}$ and $\rv_{\alpha,V}$ that determines the optimal solution to for the global omniscience. }
	       \BlankLine
           $\alphaU \leftarrow \sum_{i \in V} \frac{H(V) - H(\Set{i})}{|V| - 1}$ for the asymptotic model or $\alphaU \leftarrow \big\lceil \sum_{i \in V} \frac{H(V) - H(\Set{i})}{|V| - 1} \big\rceil$ for the non-asymptotic model\;
           Call $\PAR(H,V,\Phi)$\;
           \For{$i=2$ \emph{\KwTo} $|V|$}{
                \lIf{$\exists C \in \Qat{\alphaU}{V_i} \colon |C| > 1$ at some iteration $i$ of $\PAR$}{
                    \textbf{break} and \Return $C$ and $\rv_{\alphaComp,C}$ after the update in step~\ref{step:UpdatesParAux} of the $\PAR$ algorithm, where \label{step:alphaComp}
                        $$ \alphaComp = \min \Set{ \alpha \in \Real \colon \Fu{\alpha} (C) = \FuHat{\alpha} (C) }, $$
                        $$ \alphaComp = \min \Set{ \alpha \in \Z \colon \Fu{\alpha} (C) = \FuHat{\alpha} (C) }, $$
                    for the asymptotic and non-asymptotic models, respectively (see Remark~\ref{rem:SOPAR} for how to obtain $\alphaComp$) }
            }
            \Return $\emptyset$, $\Qat{\alpha}{V}$ and $\rv_{\alpha,V}$ at the end of the $\PAR$ algorithm\;
	   \caption{Two-stage Successive Omniscience (SO) by PAR Algorithm}
	   \end{algorithm}

\begin{remark}[Determining $\alphaComp$] \label{rem:SOPAR}
    To obtain the value of $\alphaComp$ in Lemma~\ref{lemma:SOPAR}, we just need to consider the range $[0,\alphaU]$ for that $\alphaComp \leq \alphaU$ must hold in both asymptotic and non-asymptotic models.
    For the asymptotic model, $\alphaComp = \min \Set{ \alpha \in \Real \colon \Fu{\alpha} (C) = \FuHat{\alpha} (C})$ is one of the critical points where the subsets in $C$ merge to form $C$, or the first time that $C$ appears as an intact subset in $\Qat{\alpha}{V_i}$. The reason is that: (i) for all $\alpha < \alphaComp$, $\Fu{\alpha} (C) < \FuHat{\alpha} (C)$ and therefore $C \notin \bigwedge \argmin_{\Pat \in \Pi(V_i)} \Fu{\alpha} [\Pat] = \Qat{\alpha}{V_i}$; (ii) for all $\alpha \geq \alphaComp$, $\Fu{\alpha} (C) = \FuHat{\alpha} (C)$ so that $C \subseteq C'$ for some $C' \in \Qat{\alpha}{V_i}$. So, $\alphaComp$ must create a critical value in the segmented $\Qat{\alpha}{V_i}$. Taking the least integer value that is no less than this critical value, we have $\alphaComp$ for the non-asymptotic model in Lemma~\ref{lemma:SOPAR}(b). See Example~\ref{ex:1stSO}.
\end{remark}

Also note that, according to Corollary~\ref{coro:SOSufLB}, Lemma~\ref{lemma:SOPAR} also holds for a tighter lower bound for both asymptotic and non-asymptotic models, e.g., the ones in \eqref{eq:LB}, based on which Algorithm~\ref{algo:1stSO} might return a larger (in size) complimentary subsets. See the example below.

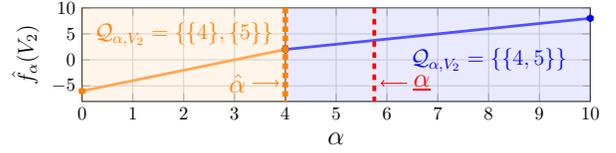
\begin{figure}[t]
	\centering
    \scalebox{0.7}{
%
%
\definecolor{mycolor1}{rgb}{1,0,1}%

\begin{tikzpicture}[
every pin/.style={rectangle,rounded corners=3pt,font=\tiny},
every pin edge/.style={<-}]

\begin{axis}[%
width=3.8in,
height=0.7in,
scale only axis,
xmin=0,
xmax=10,
xlabel={\Large $\alpha$},
xmajorgrids,
ymin=-8,
ymax=10,
ylabel={\large $\FuHat{\alpha}(V_2)$},
ymajorgrids,
legend style={at={(0.65,0.05)},anchor=south west,draw=black,fill=white,legend cell align=left}
]

\addplot [
color=blue,
solid,
line width=1.5pt,
mark=asterisk,
mark options={solid}
]
table[row sep=crcr]{
10 8\\
4 2\\
};

\addplot[area legend,solid,fill=blue!20,opacity=4.000000e-01]coordinates {
(10,10)
(10,-8)
(4,-8)
(4,10)
};
\node at (axis cs:8,0) {\textcolor{blue}{\large $\Qat{\alpha}{V_2} = \Set{\Set{4,5}}$}};

\addplot [
color=orange,
solid,
line width=1.5pt,
mark=asterisk,
mark options={solid}
]
table[row sep=crcr]{
4 2\\
0 -6 \\
};

\addplot[area legend,solid,fill=orange!20,opacity=4.000000e-01]coordinates {
(4,10)
(4,-8)
(0,-8)
(0,10)
};
\node at (axis cs:2,5) {\textcolor{orange}{\large $\Qat{\alpha}{V_2} = \Set{\Set{4},\Set{5}}$}};

\addplot [
color=red,
dashed,
line width=2pt,
]
table[row sep=crcr]{
5.75 10\\
5.75 -8\\
};
\node at (axis cs:5.75,-4.5) [pin={[pin distance = 5mm,pin edge = {red}]0:\textcolor{red}{\Large $\alphaU$}}] {};

\addplot[color=orange,dotted,line width=3pt] coordinates {
(4,10)
(4,-8)
};
\node at (axis cs:4,-4.5) [pin={[pin distance = 5mm,pin edge = {orange}]180:\textcolor{orange}{\Large $\alphaComp$}}] {};

\end{axis}
\end{tikzpicture}%
}
	\caption{For the $5$-user system in Example~\ref{ex:SOSuf}, when Algorithm~\ref{algo:1stSO} considers $\Qat{\alpha}{V_2}$ obtained at the end of $2$nd iteration of the $\PAR$ algorithm, we have the lower bound on the minimum sum-rate being $\alphaU = 5.75$ and $\Set{4,5} \in \Qat{5.75}{V_2}$ being a complimentary subset for the SO in the asymptotic model. We obtain $\alphaComp = 4$ where $\Set{4}$ and $\Set{5}$ are merged to form a subset $\Set{4,5}$ in $\Qat{\alphaComp}{V_2}$. For the $\rv_{\alpha,V_2}$ in \eqref{eq:ExQatRSO} obtained by the $\PAR$ algorithm, $\rv_{4,V_2} = \rv_{4,\Set{4,5}} = (2,0)$ is an optimal rate vector for attaining the local omniscience in $\Set{4,5}$ with the minimum sum-rate $\RACO(\Set{4,5}) = 2$. }
	\label{fig:SO2}
\end{figure}

\begin{example}[two-stage SO by Algorithm~\ref{algo:1stSO}] \label{ex:1stSO}
    We run Algorithm~\ref{algo:1stSO} on the $5$-user system in Example~\ref{ex:SOSuf} for linear ordering $\Phi = (4,5,2,3,1)$. We set $\alphaU = \sum_{i \in V} \frac{H(V) - H(\Set{i})}{|V| - 1} = 5.75$ for the asymptotic model. At the end of the $2$nd the iteration of the $\PAR$ algorithm, we have
    $$ \U{\alpha}{V_2} = \begin{cases}
                             \Set{\Set{5}} & \alpha \in [0,4],\\
                             \Set{\Set{4},\Set{5}} & \alpha \in (4,10],
                         \end{cases} $$
    where $\TU{5.75}{V_2} = \Set{4,5}$ is nonsingleton so that $C = \Set{4,5} \in \Qat{5.75}{V_2} = \Set{\Set{4,5}}$ is a complimentary subset.
    To determine the optimal rate vector for the local omniscience in $\Set{4,5}$, we use Remark~\ref{rem:SOPAR} to search the value of $\alphaComp = \min \Set{ \alpha \in \Real \colon \Fu{\alpha} (\Set{4,5}) = \FuHat{\alpha} (\Set{4,5})}$ in the segmented $\Qat{\alpha}{V_2}$ in Fig.~\ref{fig:SO2} and find that, at $\alpha = 4$, the subsets $\Set{4}$ and $\Set{5}$ are merged to $\Set{4,5}$. So, $\alphaComp = 4$. Note, $\alphaComp = 4$ is also an critical point in the segmented $\Qat{\alpha}{V_2}$. See Fig~\ref{fig:SO2}.
    Considering the segmented rate vector
    \begin{equation} \label{eq:ExQatRSO}
            \rv_{\alpha,V_2} = \rv_{\alpha,\Set{4,5}} =  \begin{cases}
                                                            (\alpha-2,\alpha-4) & \alpha \in [0,4], \\
                                                            (\alpha-2, 0) & \alpha \in (4,10]
                                                        \end{cases} \\
    \end{equation}
    determined by the $\PAR$ algorithm, we have $\rv_{4,\Set{4,5}} = (2,0)$ being the optimal rate vector that attains the local omniscience in $\Set{4,5}$ with the minimum sum-rate $\RACO(\Set{4,5}) = 2$.

    For the non-asymptotic model, we have the lower bound $\alphaU = \lceil \sum_{i \in V} \frac{H(V) - H(\Set{i})}{|V| - 1} \rceil = 6$. Again, $\Set{4,5} \in \Qat{\alphaU}{V_2}$ is complimentary and $\alphaComp = \min \Set{ \alpha \in \Z \colon \Fu{\alpha} (\Set{4,5}) = \FuHat{\alpha} (\Set{4,5})} = 4$, where $\rv_{4,\Set{4,5}} = (2,0)$ is an optimal rate vector for the local omniscience in $\Set{4,5}$.

    Consider the optimal rate vector $(1,0.5,0.5,4.5,0)$ for the global omniscience obtained in \cite[Example~\ref{ex:ParAlgo}]{DingITCO2019}, we have $(1,0.5,0.5,4.5,0) = (0,0,0,2,0) + (1,0.5,0.5,2.5,0)$, which means that, by letting users transmit at rates $(1,0.5,0.5,2.5,0)$ after reaching the local omniscience in $\Set{4,5}$, the global omniscience is still attained with the minimum sum-rate $\RACO(V) = 6.5$.
    Alternatively, it also means that the optimal rate vector $(1,0.5,0.5,4.5,0)$ can be implemented in a successive manner so that the local omniscience in $\Set{4,5}$ can be attained first. In fact, the necessary and sufficient condition for a nonsingleton $\XComp \subsetneq V$ to be complimentary in \cite[Theorems 4.2 and 5.2]{ChanSuccessiveIT} are derived in terms of the existence of the successive optimal rates: $\XComp$ is complimentary if and only if there exists an optimal rate vector $\rv_V \in \RRACO(V)$ such that $\rv_{V}^{(1)} + \rv_{V}^{(2)} = \rv_V$ with $\rv_{\XComp}^{(1)} \in \RRACO(\XComp)$ and $r_i^{(1)} = 0 $ for all $i \in V \setminus \XComp$.
    Likewise, the optimal rate vector $(0,1,1,5,0)$ for the non-asymptotic model can be implemented in a successive manner as $(0,1,1,5,0) = (0,0,0,2,0) + (0,1,1,3,0)$, where the local omniscience in $\Set{4,5}$ can be attained first.

    It should be note that Lemma~\ref{lemma:SOPAR} (a) and (b) holds for any lower bound $\alphaU$.
    The following example shows that a larger complimentary subset can be found with a tighter lower bound. With $\alphaU = 6.25$, we apply Algorithm~\ref{algo:1stSO} to the asymptotic model for another linear ordering $\Phi = (5,1,4,2,3)$. We have $\TU{6.25}{V_i}$ remains singleton until the end of $3$rd iteration in the PAR algorithm, where $\TU{6.25}{V_3} = \Set{1,4,5}$ and the segmented rate vector $\rv_{\alpha,V_3}$ and partition $\Qat{\alpha}{V_3}$ are respectively
    \begin{equation}
        \begin{aligned}
            & \rv_{\alpha,\Set{1,4,5}} = \begin{cases}
                                    (\alpha-5, \alpha-2, \alpha-4) & \alpha \in [0,4], \\
                                    (\alpha-5 ,  2, \alpha-4) & \alpha \in (4,6], \\
                                    (\alpha-5 , 8-\alpha, \alpha-4) & \alpha \in (6,7], \\
                                    (2 , 1, \alpha-4) & \alpha \in (7,10], \\
                                \end{cases}\\
            & \Qat{\alpha}{\Set{1,4,5}} = \begin{cases}
                                    \Set{\Set{1},\Set{4},\Set{5}} & \alpha \in [0,4], \\
                                    \Set{\Set{4,5},\Set{1}} & \alpha \in (4,6], \\
                                    \Set{\Set{1,4,5}} & \alpha \in (6,10].
                                \end{cases}
        \end{aligned} \nonumber
    \end{equation}
    In this case, we have the nonsingleton $\TU{6.25}{V_3} = \Set{1,4,5} \in \Qat{6.25}{\Set{1,4,5}}$, $\alphaComp = \min \Set{ \alpha \in \Real \colon \Fu{\alpha} (\Set{1,4,5}) = \FuHat{\alpha} (\Set{1,4,5})} = 6$ and $\rv_{6,\Set{1,4,5}} = (1,2,2) \in \RRACO(\Set{1,4,5})$ being the optimal rate vector that attains local omniscience in $\Set{1,4,5}$ with the minimum sum-rate $\RACO(\Set{1,4,5}) = 5$.
\end{example}

\subsection{Recursive Two-stage SO}
\label{subsec:Rec1stSO}

The two-stage SO can be recursively implemented.
After the local omniscience in the complimentary subset $\XComp$ is attained, the users in $i \in \XComp$ can be treated as a super-user $\TXComp$ that observes the source $\RZ{\TXComp} = \RZ{\XComp}$. For all $i \in V \setminus \XComp$, we need to update $\RZ{i} \leftarrow (\RZ{i},\Gamma)$ with $\Gamma$ being the broadcasts overheard by user $i$ when the users in $\XComp$ are attaining local omniscience.
The new system $V' = \Set{\TXComp} \sqcup \Set{i \colon i \in V \setminus \XComp}$ poses a new omniscience problem, which can also be solved successively.
This recursive method results in a multi-stage SO. An example can be found in \cite{Ding2017ISIT}.
We show in the next section that, without incurring any transmission, a multi-stage SO can be outlined at the end of the $\PAR$ algorithm.

\section{Multi-stage Successive Omniscience}
\label{sec:MstSO}

We first derive the necessary conditions for a multi-stage SO to be achievable and then use the partition $\Qat{\alpha}{V}$ and rate vector $\rv_{\alpha,V}$ at the end of the $\PAR$ algorithm to outline this multi-stage SO for both asymptotic and non-asymptotic models.

\subsection{Achievability}

For $K \leq |V| - 1$, let $\Set{(\XCompk{k},\rv_{V}^{(k)}) \colon k \in \Set{1,\dotsc,K}}$ denote the \emph{$K$-stage SO}, where all $\XCompk{k}$ are nonsingleton subsets of $V$. The following proposition states the achievability of the $K$-stage SO.

\begin{proposition}\label{Prop:MstSO}
    In the asymptotic model, a $K$-stage SO $\Set{(\XCompk{k},\rv_{V}^{(k)}) \colon k \in \Set{1,\dotsc,K}}$ is achievable if $\XCompk{k}$ for all $k \in \Set{1,\dotsc,K}$ is complimentary and forms a set sequence/chain
    \begin{equation}\label{eq:MSOSetChain}
                \emptyset \subsetneq \XCompk{1} \subsetneq \XCompk{2} \subsetneq \dotsc \subsetneq \XCompk{K} = V.
    \end{equation}
    For all $k \in \Set{1,\dotsc,K-1}$,
    \begin{enumerate}[(a)]
      \item $\rv_{\XCompk{k}}^{(k)} \in \RRCO(\XCompk{k})$ and $r_{i}^{(k)} = 0$ for all $i \in V \setminus \XCompk{k}$;
      \item $r^{(k+1)}(\XCompk{k}) - r^{(k)}(\XCompk{k}) \geq 0$,
    \end{enumerate}
    and $\rv_{V}^{(K)} \in \RRACO(V)$.
    In the non-asymptotic model, a $K$-stage SO is achievable if the above conditions hold for $\rv_{V}^{(K)} \in \RRNCO(V)$ and $\rv_{\XCompk{k}} \in \RRCO(\XCompk{k}) \cap \Z^{|V|}$ for all $k \in \Set{1,\dotsc,K-1}$.
\end{proposition}
\begin{IEEEproof}
    The nesting subset chain \eqref{eq:MSOSetChain} and the nondecreasing sum-rate $r^{(k)}(\XCompk{k})$ in $k$ in (b) are the necessary conditions for local omniscience in $\XCompk{k}$ to be implemented subsequently. The condition in (a) ensures the local omniscience in $\XCompk{k}$ is attained by $\rv_{\XCompk{k}}$ while the rest of users are overhearing. The optimality of the final rate vector is ensured by $\rv_{V}^{(K)} \in \RRACO(V)$. By restricting these conditions hold for the integer-valued rate vectors $\rv_{V}^{(k)}$ for all $k \in \Set{1,\dotsc,K}$, the proposition applies to the non-asymptotic model.
\end{IEEEproof}

\subsection{Asymptotic Model}

We propose Algorithm~\ref{algo:MstSOACO} that uses $\Patp{j}$s in the PSP and the corresponding rate vector $\rv_{\alphap{j},V}$ to build a $p$-stage SO that iteratively attains the local omniscience in all nonsingleton subsets in each partition $\Patp{j}$.
The achievability of this $p$-stage SO is stated below. The proof in Appendix~\ref{app:coro:MstSOACO} is based on the monotonic sum-rate for all $\X \in \Patp{j}$ in \cite[Lemma~\ref{lemma:EssProp}(c)]{DingITCO2019}.

        \begin{algorithm} [t]
	       \label{algo:MstSOACO}
	       \small
	       \SetAlgoLined
	       \SetKwInOut{Input}{input}\SetKwInOut{Output}{output}
	       \SetKwFor{For}{for}{do}{endfor}
            \SetKwRepeat{Repeat}{repeat}{until}
            \SetKwIF{If}{ElseIf}{Else}{if}{then}{else if}{else}{endif}
	       \BlankLine
           \Input{$H$, $V$ and $\Phi$.}
	       \Output{an achievable $p$-stage SO $\Set{(\XCompSetk{k},\rv_{V}^{(k)}) \colon k \in \Set{1,\dotsc,p}}$. }
	       \BlankLine
            Call $\PAR(H,V,\Phi)$ to obtain the segmented $\Qat{\alpha}{V}$ and $\rv_{\alpha,V}$ for all $\alpha$\;
            Initiate $\rv_V^{(0)} \leftarrow (0,\dotsc,0)$\;
            \For{$k=1$ \emph{\KwTo} $p$}{
                $\rv_V^{(k)} \leftarrow \rv_V^{(k-1)}$\;
                $\XCompSetk{k} \leftarrow \Set{C \in \Patp{p-k} \colon |C| > 1}$\label{step:MstSOXCompSetk}\;
                \ForEach{$C \in \XCompSetk{k}$}{
                    \lIf{$C \neq \ASet{C}{\Patp{p-k+1}}$}{
                        for each $C' \in \ASet{C}{\Patp{p-k+1}}$ randomly select user $i \in C'$ and let $\Delta r \leftarrow r_{\alphap{p-k+1}}(C') - r^{(k)}(C')$ and $r_i^{(k)} \leftarrow r_i^{(k)} + \Delta r$\label{step:MstSODecomp}
                        }
                }
            }
            \Return $\XCompSetk{k}$ and $\rv_{V}^{(k)}$ for all $k \in \Set{1,\dotsc,p}$ \;
	   \caption{Multi-stage Successive Omniscience (SO) by the $\PAR$ Algorithm for the Asymptotic Model}
	   \end{algorithm}

\begin{corollary} \label{coro:MstSOACO}
    For the asymptotic model, all $\XCompSetk{k}$ and $\rv_{V}^{(k)}$ at the end of Algorithm~\ref{algo:MstSOACO} constitute an achievable $p$-stage SO $\Set{(\XCompSetk{k},\rv_{V}^{(k)}) \colon k \in \Set{1,\dotsc,p}}$, where $\rv_V^{(p)} \in \RACO(V)$,
    $$ \emptyset \subsetneq \TXCompSetk{1} \subsetneq \dotsc \subsetneq \TXCompSetk{p} = V $$
    and, for all $k \in \Set{1,\dotsc,p}$, all $C \in \XCompSetk{k}$ are complimentary. For each $C \in \XCompSetk{k}$, the local omniscience in each $C$ is attained by an optimal rate vector $\rv_{C}^{(k)} \in \RACO(C)$. \hfill \IEEEQED
\end{corollary}

\begin{remark} \label{rem:MstSO}
    We remark the followings about Algorithm~\ref{algo:MstSOACO}.
    \begin{enumerate}[(a)]
      \item The output $\rv_V^{(k)}$ is nondecreasing in $k$, i.e., $\rv_V^{(k-1)} \geq \rv_V^{(k)}$ for all $k \in \Set{2,\dotsc,p}$.\footnote{We say $\rv_V \geq \rv'_V$ if $r_i \geq r'_i$ for all $i \in V$ with at least one of these inequalities holding strictly.}
          Therefore, $\rv_V^{(k)}$ is not necessarily the same as $\rv_{\alphap{p-k+1},V}$ in that $\rv_{\alpha,V}$ returned by the $\PAR$ algorithm is not monotonic in general, e.g., $r_{\alpha,3}$ in \eqref{eq:RpV5Aux} is not nondecreasing in $\alpha$.
      \item Unlike Proposition~\ref{Prop:MstSO}, where only one complimentary subset is chosen each time, Algorithm~\ref{algo:MstSOACO} allows more than one complimentary subset to attain local omniscience at each stage. Since all $C \in \XCompSetk{k}$ are disjoint, the local omniscience in step~\ref{step:MstSODecomp} can be attained simultaneously if the broadcasts between subsetes do not cause interference, e.g., via orthogonal wireless channels in CCDE.\footnote{For example, if $\Patp{p-1} = \Set{\Set{1,2},\Set{3},\Set{4,5}}$ at the $1$st stage of SO, the users $1$, $2$, $4$ and $5$ can transmit at the same time to attain the local omniscience in $\Set{1,2}$ and $\Set{4,5}$, respectively. The purpose of the decomposition $\ASet{C}{\Pat}$ in step~\ref{step:MstSODecomp} is to search all users/super-users that are supposed to take part in the local omniscience in $C \in \Patp{j}$. See Example~\ref{ex:MstSOACO}. }
      \item $\Delta r$ in \eqref{eq:DeltaR} is interpreted as, in addition to the rates for attaining the local omniscience in $C'$, how many transmissions is required from the super-user $C'$ for attaining the local omniscience in $C$. Since all users in $C'$ have recovered $\RZ{C'}$ in previous stages, $\Delta r$ can be assigned to any one of them. Apart from the random selection in step~\ref{step:MstSODecomp}, we can moderate $\Delta r$ to the users $i$ with the lowest $r_i^{(k)}$ for improving fairness. See Example~\ref{ex:MstSOACO}.
    \end{enumerate}
\end{remark}

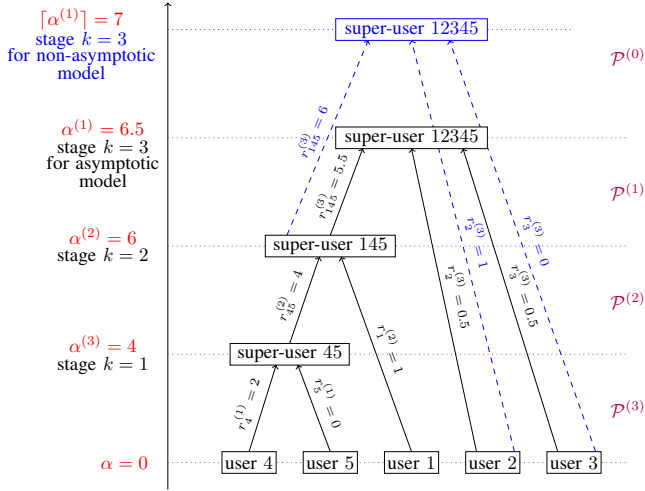
\begin{figure}[t]
	\centering
    \scalebox{0.72}{\begin{tikzpicture}

\draw [->] (0,-0.5)--(0,8.5);

\node at (-0.8,0) {\textcolor{red}{$\alpha = 0$}};
\draw [dotted , red](0,0) -- (8.5,0);
\node at (8.5,1) {\textcolor{purple}{$\mathcal{P}^{(3)}$}};

\draw [fill = white] (1,0.2) rectangle (2,-0.2);
\node at (1.5,0) {user $4$};

\draw [fill = white] (2.5,0.2) rectangle (3.5,-0.2);
\node at (3,0) {user $5$};

\draw [fill = white] (4,0.2) rectangle (5,-0.2);
\node at (4.5,0) {user $1$};

\draw [fill = white] (5.5,0.2) rectangle (6.5,-0.2);
\node at (6,0) {user $2$};

\draw [fill = white] (7,0.2) rectangle (8,-0.2);
\node at (7.5,0) {user $3$};

\node at (-1.2,2.2) {\textcolor{red}{$\alpha^{(3)} = 4$}};
\node at (-1.2,1.8) {stage $k=1$};
\draw [dotted, red](0,2) -- (8.5,2);
\node at (8.5,3) {\textcolor{purple}{$\mathcal{P}^{(2)}$}};

\draw [fill = white] (1.15,2.2) rectangle (3.35,1.8);
\node at (2.25,2) {super-user $45$};
\draw [->] (1.5,0.2)--node [sloped, above] {\scriptsize $r_4^{(1)} = 2$}(2,1.8);
\draw [->] (3,0.2)--node [sloped, above] {\scriptsize $r_5^{(1)} = 0$}(2.4,1.8);

\node at (-1.2,4.2) {\textcolor{red}{$\alpha^{(2)} = 6$}};
\node at (-1.2,3.8) {stage $k=2$};
\draw [dotted, red](0,4) -- (8.5,4);
\node at (8.5,5) {\textcolor{purple}{$\mathcal{P}^{(1)}$}};

\draw [fill = white] (1.8,4.2) rectangle (4.2,3.8);
\node at (3,4) {super-user $145$};
\draw [->] (2.25,2.2)--node [sloped, above] {\scriptsize $r_{45}^{(2)} = 4$}(2.8,3.8);
\draw [->] (4.5,0.2)--node [sloped, above] {\scriptsize $r_{1}^{(2)} = 1$}(3.2,3.8);

\node at (-1.2,6.2) {\textcolor{red}{$\alpha^{(1)} = 6.5$}};
\node at (-1.2,5.8) {stage $k=3$};
\node at (-1.2,5.5) {for asymptotic};
\node at (-1.2,5.2) {model};
\draw [dotted, red](0,6) -- (8.5,6);
\node at (8.5,7.5) {\textcolor{purple}{$\mathcal{P}^{(0)}$}};

\draw [->] (3,4.2)--node [sloped, above] {\scriptsize $r_{145}^{(3)} = 5.5$}(3.6,5.8);
\draw [->] (5.7,0.2)--node [sloped, above] {\scriptsize $r_{2}^{(3)} = 0.5$}(4.5,5.8);
\draw [->] (7.2,0.2)--node [sloped, above] {\scriptsize $r_{3}^{(3)} = 0.5$}(5.45,5.8);

\node at (-1.6,8.2) {\textcolor{red}{$\lceil \alpha^{(1)} \rceil = 7$}};
\node at (-1.6,7.8) {\textcolor{blue}{stage $k=3$}};
\node at (-1.6,7.5) {\textcolor{blue}{for non-asymptotic}};
\node at (-1.6,7.2) {\textcolor{blue}{model}};
\draw [dotted, blue](0,8) -- (7.5,8);

\draw [->,dashed,blue] (2.2,4.2)--node [sloped, above] {\scriptsize \textcolor{blue}{$r_{145}^{(3)} = 6$}}(3.7,7.8);
\draw [->,dashed,blue] (6.4,0.2)--node [sloped, above] {\scriptsize \textcolor{blue}{$r_{2}^{(3)} = 1$}}(4.5,7.8);
\draw [->,dashed,blue] (7.9,0.2)--node [sloped, above] {\scriptsize \textcolor{blue}{$r_{3}^{(3)} = 0$}}(5.2,7.8);

\draw [blue,fill = white] (3.1,8.2) rectangle (5.9,7.8);
\node at (4.5,8) {\textcolor{blue}{super-user $12345$}};
\draw [fill = white] (3.1,6.2) rectangle (5.9,5.8);
\node at (4.5,6) {super-user $12345$};

\end{tikzpicture}}
	\caption{The $3$-stage agglomerative SO tree outlined in Example~\ref{ex:MstSOACO} by applying Algorithm~\ref{algo:MstSOACO} to the $5$-user system for the asymptotic model in Example~\ref{ex:SOSuf}. Here, the users/super-users at each stage $k$ correspond to a $\Patp{j}$ in the PSP of $V$, which characterizes the segmented partition $\Qat{\alpha}{V}$. The rates $r_i^{(k)}$ are determined by the segmented rate vector $\rv_{\alpha,V}$ in \eqref{eq:RpV5Aux}. The $3$-stage SO for the non-asymptotic model determined in Example~\ref{ex:MstSONCO} by Algorithm~\ref{algo:MstSONCO} is also shown, which only differs from the one for the asymptotic model at the last stage $k = 3$, where the super-user $145$ and users $2$ and $3$ merges at $\alpha = 7$ instead of $\alpha = 6.5$. }
	\label{fig:MSOTree}
\end{figure}

Intuitively, the $p$-stage SO $\Set{(\XCompSetk{k},\rv_{V}^{(k)}) \colon k \in \Set{1,\dotsc,p}}$ results in an agglomerative SO tree that converges to the global omniscience. See Fig.~\ref{fig:MSOTree}. This SO tree is exactly the hierarchical clustering result determined by the PSP of $V$ as described in \cite[Section~\ref{sec:Clustering}]{DingITCO2019}.
This bottom-up approach can also be considered as an opposite process of the divide-and-conquer algorithm in \cite{MiloDivConq2011}, where the ground set $V$ is recursively split into subsets until the optimal rates $r_{\RACO(V),i}$ are determined for all users $i \in V$. However, the complexity is much reduced: while the complexity of divide-and-conquer algorithm \cite{MiloDivConq2011} is $O(|V|^3 \cdot \SFM(|V|))$, Algorithm~\ref{algo:MstSOACO} completes in $O(|V| \cdot \SFM(|V|))$ time.\footnote{We neglect the computations after step 2 of Algorithm~\ref{algo:MstSOACO} because they are much less complex than the $\SFM$ algorithm. Therefore, the complexity of Algorithm~\ref{algo:MstSOACO} is the same as the $\PAR$ algorithm. }

\begin{example}\label{ex:MstSOACO}
    We apply Algorithm~\ref{algo:MstSOACO} to the $5$-user system in Example~\ref{ex:SOSuf}. The call $\PAR(H,V,(4,5,2,3,1))$ returns
    \begin{equation} \label{eq:RpV5Aux}
        \rv_{\alpha,V} = \begin{cases}
                                    (\alpha-5, \alpha-6, \alpha-6, \alpha-2, \alpha-4) & \alpha \in [0,4] ,\\
                                    (\alpha-5, \alpha-6, \alpha-6, \alpha-2, 0) & \alpha \in (4,6] ,\\
                                    (1, \alpha-6, \alpha-6, \alpha-2, 0) & \alpha \in (6,6.5] ,\\
                                    (14-2\alpha, \alpha-6, \alpha-6, \alpha-2, 0) & \alpha \in (6.5,7] ,\\
                                    (0, \alpha-6, 8-\alpha, \alpha-2, 0) & \alpha \in (7,8] ,\\
                                    (0, 2, 0, \alpha-2, 0) & \alpha \in (8,10] ,
                                \end{cases}
    \end{equation}
    and $\Qat{\alpha}{V}$ characterized by the PSP: $\Patp{3} = \Set{\Set{1},\dotsc,\Set{5}}$, $\Patp{2} = \Set{\Set{4,5},\Set{1},\Set{2},\Set{3}}$ and $\Patp{1} = \Set{\Set{1,4,5},\Set{2},\Set{3}}$ with $p = 3$ critical values $\alphap{3} = 4$, $\alphap{2} = 6$ and $\alphap{1} =6.5$. Let $\rv_V^{(0)} = (0,\dotsc,0)$.

    For $k = 1$, first assign $\rv_V^{(1)} = (0,\dotsc, 0)$. We get $\XCompSetk{1} = \Set{C \in \Patp{2} \colon |C| > 1} = \Set{\Set{4,5}}$ such that $\ASet{\Set{4,5}}{\Patp{3}} = \Set{\Set{4},\Set{5}} \neq \Set{4,5}$. This means local omniscience has not attained in $\Set{4,5}$ before. We then assign rates in step~\ref{step:MstSODecomp} as $r_{4}^{(1)} = r_{\alphap{3},4} = 2$ and $r_{5}^{(1)} = r_{\alphap{3},5} = 0$ so that $\rv_V^{(1)} = (0,0,0,2,0)$.

    For $k = 2$, assign $\rv_V^{(2)} = \rv_V^{(1)} = (0,0,0,2,0)$ and get $\XCompSetk{2} = \Set{C \in \Patp{1} \colon |C| > 1} = \Set{\Set{1,4,5}}$, where $\ASet{\Set{1,4,5}}{\Patp{2}} = \Set{\Set{1},\Set{4,5}} \neq \Set{1,4,5}$. By treating $C' = \Set{4,5}$ as a super-user, we have $r_{\alphap{2}}(\Set{4,5}) = 4$ so that $\Delta r = r_{\alphap{2}}(\Set{4,5}) - r^{(2)}(\Set{4,5}) = 2$. This means in addition to $\rv_V^{(1)} = (0,0,0,2,0)$ that attains the local omniscience in $\Set{4,5}$, users $4$ and $5$ need to transmit $2$ more times for attaining the local omniscience in $\Set{1,4,5}$. In this case, we choose user $4$ to transmit $\Delta r$ so that $r_{4}^{(2)} = 2 + 2 = 4$; For $C' = \Set{1}$ being singleton, we haven't assigned any rates to user $1$ before and therefore $r_{1}^{(2)} = r_{\alphap{2},1} = 1$. So, $\rv_V^{(2)}$ is updated to $(1,0,0,4,0)$.

    For $k = 3$, assign $\rv_V^{(3)} = \rv_V^{(2)} = (1,0,0,4,0)$ and get $\XCompSetk{1} = \Set{C \in \Patp{0} \colon |C| > 1} = \Set{\Set{1,\dotsc,5}}$, where $\ASet{\Set{1,\dotsc,5}}{\Patp{1}} = \Set{\Set{1,4,5},\Set{2},\Set{3}} \neq \Set{1,\dotsc,5}$. For the super-user formed by $C' = \Set{1,4,5}$, we have $r_{\alphap{1}}(\Set{1,4,5}) = 5.5$ and $\Delta r = r_{\alphap{1}}(\Set{1,4,5}) - r^{(3)}(\Set{1,4,5}) = 0.5$. Still choose user $4$ so that $r_{4}^{(3)} = 4 + 0.5 = 4.5$; For singletons $\Set{2}$ and $\Set{3}$, assign $r_{2}^{(3)} = r_{\alphap{1},2} = 0.5$ and $r_{3}^{(3)} = r_{\alphap{1},3} = 0.5$. So, $\rv_V^{(3)}$ is updated to $(1,0.5,0.5,4.5,0)$, which is an optimal rate vector in $\RRACO(V)$ for attaining the global omniscience. Finally, we have a $3$-stage SO $\Set{(\XCompSetk{k},\rv_{V}^{(k)}) \colon k \in \Set{1,\dotsc,3}}$ such that $ \TXCompSetk{1} \subsetneq \TXCompSetk{2} \subsetneq \TXCompSetk{3} = V $ and $\rv_{V}^{(1)} \leq \rv_{V}^{(2)} \leq \rv_{V}^{(3)}$.
    %

    While the above procedure outputs $\rv_V^{(k)} = \rv_{\alphap{p-k+1},V}$ for all $k$, we show that a fair allocation of $\Delta r$ in Remark~\ref{rem:MstSO}(c) results in a different $\rv_V^{(k)}$. If we assign $\Delta r = r_{\alphap{2}}(\Set{4,5}) - r^{(2)}(\Set{4,5}) = 2$ to user $5$ in stage $k = 2$ and $\Delta r = r_{\alphap{1}}(\Set{1,4,5}) - r^{(3)}(\Set{1,4,5}) = 0.5$ to user $1$, we have a fairer rate vector sequence $\rv_V^{(1)} = (0,0,0,2,0)$, $\rv_V^{(2)} = (1,0,0,2,2)$ and $\rv_V^{(3)} = (1.5,0.5,0.5,2,2) \in \RRACO(V)$. However, in general, this approach does not necessarily result in the fairest optimal rate vector $\RRACO(V)$ at the end of final stage $k = p$.\footnote{This is the case since, in the first stage $k = 1$, the rate vector $\rv_{C}^{(1)}$ is already an extreme point or vertex in $\RRACO(C)$ for all $C \in \XCompSetk{1}$ \cite[Theorem~27]{Ding2018IT}. Thus, Remark~\ref{rem:MstSO}(c) only attains some level of fairness in the optimal rate vector set $\RRACO(V)$. Independent from this successive approach, there are several algorithms proposed in \cite{Ding2019Fair} for searching the fairest optimal rate vector for both asymptotic and non-asymptotic models. }

    The $3$-stage SO above can be presented as the agglomerative tree diagram in Fig.~\ref{fig:MSOTree}, which provides a more intuitive interpretation of what Algorithm~\ref{algo:MstSOACO} does: attain local omniscience in all subsets merged from $\Patp{k}$ to $\Patp{k+1}$ at each $\alphap{k}$.
    In Fig.~\ref{fig:MSOTree}, increasing $\alpha$ from $0$ at the bottom, at $\alphap{3} = 4$, the value of $\Qat{\alpha}{V}$ changes from $\Patp{3} = \Set{\Set{1},\dotsc,\Set{5}}$ to $\Patp{2} = \Set{\Set{4,5},\Set{1},\Set{2},\Set{3}}$ where users $4$ and $5$ attain omniscience and merge to the super-user $45$; at $\alphap{2} = 6$, $\Qat{\alpha}{V}$ changes from $\Patp{2} = \Set{\Set{4,5},\Set{1},\Set{2},\Set{3}}$ to $\Patp{1} = \Set{\Set{1,4,5},\Set{2},\Set{3}}$ where super-users $45$ and user $1$ attain omniscience and merge to the super-user $145$; at $\alphap{1} = 6.5$, $\Qat{\alpha}{V}$ changes from $\Patp{1} = \Set{\Set{1,4,5},\Set{2},\Set{3}}$ to $\Patp{0} = \Set{\Set{1,\dotsc,5}}$ where super-users $145$, user $2$ and user $3$ attain omniscience and merge to the super-user $12345$ so that global omniscience is attained.
\end{example}

\subsection{Non-asymptotic Model}

Although all nonsingleton $C \in \Qat{\alpha}{V}$ for each integer-valued $\alpha \in \Set{0, \dotsc,\RNCO(V)}$ are complimentary based on Lemma~\ref{lemma:SOSuf}, the rate vector $\rv_{\alpha,V}$ is not necessarily nondecreasing in integer-valued $\alpha$.\footnote{An example is the rate $r_{\alpha,3}$ in \eqref{eq:RpV5Aux}, where $r_{7,3} = 1$ but $r_{9,3} = 0$ so that the monotonicity in Proposition~\ref{Prop:MstSO}(b) does not hold.}
This means that Algorithm~\ref{algo:MstSOACO} cannot be applied to the non-asymptotic model by simply running each stage $k$ of Algorithm~\ref{algo:MstSOACO} at the integer-valued critical points $\lceil \alphap{p-k+1} \rceil$.
In this section, we show that, an achievable $K$-stage SO $\Set{(\XCompk{k},\rv_{V}^{(k)}) \colon k \in \Set{1,\dotsc,K}}$ with the integer-valued $\rv_V^{(k)}$ attaining local omniscience in each complimentary subset $\XCompk{k}$ can be searched by no more than two calls of the $\PAR$ algorithm.

\begin{corollary} \label{coro:MstSONCO}
    In the non-asymptotic model, for any nonsingleton subset sequence
       $\XCompk{1} \subsetneq \dotsc \subsetneq \XCompk{K} = V$
    and the integer-valued sequence $\alphaNCOp{1} < \dotsc < \alphaNCOp{K} = \RNCO(V)$ such that $\alphaNCOp{k} \in [\alphap{j},\alphap{j-1})$ for some $j \in \Set{1,\dotsc,p}$ and $\XCompk{k} \in \Patp{j-1}$, let $\PhiNCO = (\phiNCO_1,\dotsc,\phiNCO_{|V|})$ be a linear ordering such that\footnote{The linear ordering $\PhiNCO $ satisfying \eqref{eq:PhiNCO} ensures $V_{|\XCompk{k}|} = \XCompk{k}$ for all $k \in \Set{1,\dotsc,K}$.}
    \begin{equation} \label{eq:PhiNCO}
        \phiNCO_{i} < \phiNCO_{i'}, \quad \forall \phiNCO_{i} \in \XCompk{k}, \phiNCO_{i'} \in \XCompk{k'} \setminus \XCompk{k}  \colon k < k'.
    \end{equation}
    and $\rvNCO_{\alpha,V}$ be the rate vector returned by the call $\PAR(V,H,\PhiNCO)$. For all $k \in \Set{1,\dotsc,K}$, the integer-valued $\rvNCO_{\alphaNCOp{k},\XCompk{k}}$ attains local omniscience in $\XCompk{k}$; For all $k \in \Set{1,\dotsc,K-1}$, $\rNCO_{\alphaNCOp{k+1}} (\XCompk{k}) \geq \rNCO_{\alphaNCOp{k}} (\XCompk{k})$. \hfill\IEEEQED
\end{corollary}

Corollary~\ref{coro:MstSONCO} holds the achievability of the multi-stage SO in Proposition~\ref{Prop:MstSO} for the non-asymptotic model. The proof is in Appendix~\ref{app:coro:MstSONCO}.
We propose Algorithm~\ref{algo:MstSONCO} for determining this multi-stage SO, where the purpose of steps~\ref{step:MstSOCoroStart} to \ref{step:MstSOCoroEnd} is to obtain the values of $\XCompk{k}$, $\alphaNCOp{k}$ and $\PhiNCO$ in Corollary~\ref{coro:MstSONCO}. We explain these steps as follows.
While each $\alphaNCOp{k}$ can be any integer-valued $\alpha \in [\alphap{j},\alphap{j-1})$, we choose $\alphaNCOp{k}$ to be the minimum integer in the range $[\alphap{j},\alphap{j-1})$ so that the last value is $\alphaNCOp{K} = \min \Set{\alpha \colon \alpha \in [\alphap{1},\alphap{0}) \cap \Z} = \RNCO(V)$.
Note, we could have the number of stages $K < p$ if there is no integer in some range $[\alphap{j},\alphap{j-1})$.
Once $\alphaNCOp{1}$ is determined, randomly select any nonsingleton $C \in \Patp{p-1}$ and assign to $\XCompk{1}$ and, for each $k > 1$, choose $\XCompk{k}$ as any subset $C \in \Patp{j-1}$ such that $\XCompk{k-1} \subsetneq C$ so that $\XCompk{1} \subsetneq \dotsc \subsetneq \XCompk{k-1} \subsetneq C = \XCompk{k}$ maintains for all $k$ with the last subset being $\XCompk{K} = V$ necessarily.

A linear ordering $\PhiNCO = (\phiNCO_1,\dotsc,\phiNCO_{|V|})$ satisfying \eqref{eq:PhiNCO} can be constructed by keeping each $\XCompk{k}$ the first $|\XCompk{k}|$ elements $\phiNCO_1,\dotsc,\phiNCO_{|\XCompk{k}|}$. 
To construct such a linear ordering $\PhiNCO$, let $\Set{\phiNCO_1,\dotsc,\phiNCO_{|\XCompk{1}|}} = \XCompk{1}$ and $\Set{\phiNCO_{|\XCompk{k}|+1},\dotsc, \phiNCO_{|\XCompk{k+1}|}} = \XCompk{k+1} \setminus \XCompk{k}$ for all $k \in \Set{1,\dotsc,K - 1}$.
For example, if $\XCompk{1} = \Set{3,4}$ and $\XCompk{2} = \Set{1,\dotsc,4}$, the we could have $\PhiNCO$ being $(3,4,1,2)$, $(3,4,2,1)$, $(4,3,1,2)$ or $(4,3,2,1)$, since all of them hold $V_2 = \XCompk{1} = \Set{3,4}$ and $V_4 = \XCompk{2} = \Set{1,\dotsc,4}$.

        \begin{algorithm} [t]
	       \label{algo:MstSONCO}
	       \small
	       \SetAlgoLined
	       \SetKwInOut{Input}{input}\SetKwInOut{Output}{output}
	       \SetKwFor{For}{for}{do}{endfor}
            \SetKwRepeat{Repeat}{repeat}{until}
            \SetKwIF{If}{ElseIf}{Else}{if}{then}{else if}{else}{endif}
	       \BlankLine
           \Input{$H$, $V$ and $\Phi$.}
	       \Output{an achievable $K$-stage SO $\Set{(\XCompk{k},\rv_{V}^{(k)}) \colon k \in \Set{1,\dotsc,K}}$ in the non-asymptotic model. }
	       \BlankLine
            Arbitrarily choose a linear ordering $\Phi$ and call $\PAR(H,V,\Phi)$ to obtain the segmented $\Qat{\alpha}{V}$ and $\rv_{\alpha,V}$ for all $\alpha$\;
            $k \leftarrow 1$ and $\XCompk{0} \leftarrow \emptyset$\label{step:MstSOPAR1}\;
            \For{$j = p$ \emph{decreasing from} $p$ \emph{\KwTo} $1$}{ \label{step:MstSOCoroStart}
                    \If{$[\alphap{j},\alphap{j-1}) \cap \Z \neq \emptyset$}{
                            $\alphaNCOp{k} \leftarrow \min \Set{\alpha \colon \alpha \in [\alphap{j},\alphap{j-1}) \cap \Z}$\;
                            $\XCompk{k} \leftarrow  C$, where $C \in \Patp{j-1}$ such that $\XCompk{k - 1} \subsetneq C$\;
                            $k \leftarrow k + 1$\;
                            $(\phiNCO_{|\XCompk{k-1}|+1},\dotsc, \phiNCO_{|\XCompk{k}|}) \leftarrow \XCompk{k} \setminus \XCompk{k-1}$;}
                    } \label{step:MstSOCoroEnd}
            $(\Qat{\alpha}{V}, \rvNCO_{\alpha,V}) \leftarrow \PAR(H,V,\PhiNCO)$\label{step:MstSOPAR2}\;
            Initiate $\rv_V^{(0)} \leftarrow (0,\dotsc,0)$\label{step:MstSORateStart}\;
            \For{$k=1$ \emph{\KwTo} $K$}{
                $\rv_V^{(k)} \leftarrow \rv_V^{(k-1)}$\;
                Randomly select user $i \in \XCompk{k - 1}$ and let $\Delta r \leftarrow \rNCO_{\alphaNCOp{k}}(\XCompk{k - 1}) - r^{(k)}(\XCompk{k - 1})$ and $r_i^{(k)} \leftarrow r_i^{(k)} + \Delta r$\;
                \lForEach{$i \in \XCompk{k} \setminus \XCompk{k-1}$}{let user $i$ transmit at rate $\rvNCO_{\alphaNCOp{k},i}$ and $r_i^{(k)} \leftarrow \rvNCO_{\alphaNCOp{k},i}$}
                }
            \Return $\XCompSetk{k}$ and $\rv_{V}^{(k)}$ for all $k \in \Set{1,\dotsc,K}$\label{step:MstSORateEnd}\;
	   \caption{Multi-stage Successive Omniscience (SO) by the $\PAR$ Algorithm for the Non-Asymptotic Model}
	   \end{algorithm}

\begin{example}\label{ex:MstSONCO}
    We apply Algorithm~\ref{algo:MstSONCO} to the $5$-user system in Example~\ref{ex:SOSuf}. The call $\PAR(H,V,(4,5,2,3,1))$ returns the same $\Qat{\alpha}{V}$ and $\rv_{\alpha,V}$ as in Example~\ref{ex:MstSOACO}. For the three critical points $\alphap{3} = 4$, $\alphap{2} = 6$ and $\alphap{1} =6.5$, we need to search the integer valued $\alpha$ for three regions $[4,6)$, $[6,6.5)$ and $[6.5,10)$.

    For region $[\alphap{3},\alphap{2}) = [4,6)$, we have $\alphaNCOp{1} = \min \Set{\alpha \colon \alpha \in [\alphap{3},\alphap{2}) \cap \Z} = 4$, where we assign $\XCompk{1} = \Set{4,5}$, since it is the only nonsingleton subset in $\Patp{2} = \Set{\Set{4,5},\Set{1},\Set{2},\Set{3}}$, and $(\phiNCO_1,\phiNCO_2) = (4,5)$.
    For region $[\alphap{2},\alphap{1}) = [6,6.5)$, we have $\alphaNCOp{2} = \min \Set{\alpha \colon \alpha \in [\alphap{2},\alphap{1}) \cap \Z} = 6$. We assign $\XCompk{2} = \Set{1,4,5}$ since $\Set{1,4,5} \in \Patp{1} = \Set{\Set{1,4,5},\Set{2},\Set{3}}$ and $\XCompk{1} \subsetneq \Set{1,4,5}$. We then add user $\XCompk{2} \setminus \XCompk{1} = 1$ to the linear ordering by assigning $(\phiNCO_{|\XCompk{1}|+1},\dotsc, \phiNCO_{|\XCompk{2}|}) = \phiNCO_3 = \XCompk{2} \setminus \XCompk{1} = 1$.
    For region $[\alphap{1},\alphap{0}) = [6.5,10)$, we have $\alphaNCOp{3} = \min \Set{\alpha \colon \alpha \in [\alphap{1},\alphap{0}) \cap \Z} = 7 = \RNCO(V)$. We assign $\XCompk{3} = \Set{1,\dotsc,5} = V$, where $\XCompk{2} \subsetneq \XCompk{3}$, and $(\phiNCO_{|\XCompk{2}|+1},\dotsc, \phiNCO_{|\XCompk{3}|}) = (\phiNCO_4,\phiNCO_5) = \XCompk{3} \setminus \XCompk{2} = (2,3)$ to finish constructing the linear ordering $\PhiNCO = (4,5,1,2,3)$.

    We run the $\PAR$ algorithm again with $\PhiNCO$ and get a new rate vector
    \begin{equation} \nonumber
            \rvNCO_{\alpha,V} = \begin{cases}
                                    (\alpha-5, \alpha-6, \alpha-6, \alpha-2, \alpha-4) & \alpha \in [0,4] ,\\
                                    (\alpha-5, \alpha-6, \alpha-6, \alpha-2, 0) & \alpha \in (4,6] ,\\
                                    (\alpha-5, \alpha-6, \alpha-6, \alpha-2, 0) & \alpha \in (6,6.5] ,\\
                                    (1, \alpha-6, 7-\alpha, \alpha-2, 0) & \alpha \in (6.5,7] ,\\
                                    (1, 1, 0, \alpha-2, 0) & \alpha \in (7,10].
                                \end{cases}
    \end{equation}
    One can verify that $\rNCO_{\alpha}(\XCompk{1}) = \rNCO_{\alpha}(\Set{4,5}) = \FuHat{\alpha}({4,5})$ and $\rNCO_{\alpha}(\XCompk{2}) = \rNCO_{\alpha}(\Set{1,4,5}) = \FuHat{\alpha}({1,4,5})$ for all $\alpha$ so that $\rNCO_{\alpha}(\XCompk{1})$ and $\rNCO_{\alpha}(\XCompk{2})$ is strictly increasing in $\alpha$ based on \cite[Lemma~\ref{lemma:EssProp}(c)]{DingITCO2019}. This holds the monotonicity in Proposition~\ref{Prop:MstSO}(b).

    We then initiate $\rv_V^{(0)} = (0,\dotsc,0)$ and assign the rates $\rv_V^{(k)}$ for $k \in \Set{1,\dotsc,3}$ as follows. For $k = 1$, $\alphaNCOp{1} = 4$, $\XCompk{0} = \emptyset$ and $\XCompk{1} = \Set{4,5}$. We assign $r_4^{(1)} = \rvNCO_{\alphaNCOp{1},4} = 2$ and $r_5^{(1)} = \rvNCO_{\alphaNCOp{1},5} = 0$ so that $\rv_V^{(1)} = (0,0,0,2,0)$;

    For $k = 2$ and $\alphaNCOp{2} = 6$, we choose user $4 \in \XCompk{1} = \Set{4,5}$ to transmit the rate $\Delta r = \rNCO_{\alphaNCOp{2}}(\XCompk{1}) - r^{(2)}(\XCompk{1}) = \rNCO_{6}(\Set{4,5}) - r^{(2)}(\Set{4,5}) = 4 - 2 = 2$ and update $r_4^{(2)}$ to $4$. For $\XCompk{2} \setminus \XCompk{1} = \Set{1}$, we assign user $1$ the rate $r_1^{(2)} = \rvNCO_{\alphaNCOp{2},1} = 1$ so that $\rv_V^{(2)} = (1,0,0,4,0)$;

    For $k = 3$, $\alphaNCOp{3} = 7$ and $\XCompk{3} = \Set{1,\dotsc,5}$, we choose user $4 \in \XCompk{2} = \Set{1,4,5}$ to transmit the rate $\Delta r = \rNCO_{\alphaNCOp{3}}(\XCompk{2}) - r^{(3)}(\XCompk{2}) = \rNCO_{7}(\Set{1,4,5}) - r^{(3)}(\Set{1,4,5}) = 6 - 5 = 1$ and update $r_4^{(3)}$ to $5$. For $\XCompk{3} \setminus \XCompk{2} = \Set{2,3}$, we assign user $2$  the rate $r_2^{(3)} = \rvNCO_{\alphaNCOp{3},2} = 1$ and user $1$ the rate $r_3^{(3)} = \rvNCO_{\alphaNCOp{3},3} = 0$. Finally, we have $\rv_V^{(3)} = (1,1,0,5,0) \in \RRNCO(V)$ being an optimal rate vector attains the global omniscience and the $3$-stage SO $\Set{(\XCompk{k},\rv_{V}^{(k)}) \colon k \in \Set{1,\dotsc,3}}$ being outlined as the agglomerative tree in Fig.~\ref{fig:MSOTree}, which only differs from the $3$-stage SO determined in Example~\ref{ex:MstSOACO} in the last stage $k = 3$.

    Alternatively, we can assign $\Delta r$ to the users $i$ with the lowest $r_i^{(k)}$ at stage $k = 2$ and $k = 3$, as in Remark~\ref{rem:MstSO}(c), so that we have the fairer rates $\rv_V^{(1)} = (0,0,0,2,0)$, $\rv_V^{(2)} = (1,0,0,2,2)$ and $\rv_V^{(3)} = (2,1,0,2,2) \in \RRNCO(V)$ instead.
\end{example}

Note that, we could have $\rv_{\alphaNCOp{k},V}$ returned by the first call of the $\PAR$ algorithm also achievable in the $K$-stage SO. For example, the rate vector obtained in \eqref{eq:RpV5Aux} satisfies Proposition~\ref{Prop:MstSO}(b) for $\XCompk{1} = \Set{4,5}$, $\XCompk{2} = \Set{1,4,5}$ and $\XCompk{3} = \Set{1,\dotsc,5}$. In this case, without calling the $\PAR$ algorithm, we can run steps~\ref{step:MstSORateStart} to \ref{step:MstSORateEnd} directly based on $\rv_{\alphaNCOp{k},V}$.
Therefore, an achievable $K$-stage SO for the non-asymptotic model can be determined by Algorithm~\ref{algo:MstSONCO} by no more than two calls of the $\PAR$ algorithm in $O(|V| \cdot \SFM(|V|))$ time.\footnote{These two calls output the same $\Qat{\alpha}{V}$, which is independent of the input linear ordering.}

\section{Conclusion}

We used the outputs of the $\PAR$ algorithm proposed in Part I \cite{DingITCO2019} to efficiently solve the two-stage SO and multi-stage SO problems for both asymptotic and non-asymptotic models.
We proved that a lower bound $\alphaU$ on the minimum sum-rate for the global omniscience problem is sufficient to determine the existence of a complimentary user subset for SO. For the two-stage SO, when $\alphaU$ is applied to the segmented partition $\Qat{\alpha}{V_i}$ at the end of each iteration $i$ of $\PAR$, a complimentary user subset can be extracted from $\Qat{\alphaU}{V_i}$ and the corresponding local omniscience achievable rate vector is determined by the rate vector $\rv_{\alpha,V_i}$.
The further study showed that, based on the segmented partition $\Qat{\alpha}{V}$ and rate vector $\rv_{\alpha,V}$ obtained at the end of the $\PAR$ algorithm, an achievable $p$-stage SO $\Set{(\XCompSetk{k},\rv_V^{(k)}) \colon k \in \Set{1,\dotsc,p} }$ for the asymptotic model and an achievable $K$-stage SO $\Set{(\XCompk{k},\rv_V^{(k)}) \colon k \in \Set{1,\dotsc,K}}$ for the non-asymptotic model can be both determined in $O(|V| \cdot \SFM(|V|))$ time.

As the extension of the work in this paper, it is worth studying how the agglomerative multi-stage SO tree (e.g., Fig.~\ref{fig:MSOTree}) relates to the agglomerative clustering approach in \cite{AggloClust2018ISIT}. It is also of interest to see how the results on the non-asymptotic model in this paper can be applied to practical CCDE systems, e.g., apart from random linear network coding (RLNC) \cite{Ho2006RLNC} in the recursive two-stage SO in Section~\ref{subsec:Rec1stSO}.

\appendices

\section{Proof of Corollary~\ref{coro:SOEqv}}
\label{app:coro:SOEqv}

Consider the asymptotic model first. Based on Theorem~\ref{theo:SOIff}, $ \RACO(\XComp) \leq \RACO(V) - H(V) + H(\XComp) = \Fu{\RACO(V)}(\XComp) $ is the necessary and sufficient condition for $\XComp$ to be complimentary.
We also have $\RACO(\XComp) \geq \sum_{C \in \Pat} \frac{H(\XComp) - H(C)}{|\Pat| - 1}, \forall \Pat \in \Pi(\XComp) \colon |\Pat| > 1$. So, $\sum_{C \in \Pat} \frac{H(\XComp) - H(C)}{|\Pat| - 1} \leq \RACO(V) - H(V) + H(\XComp), \forall \Pat \in \Pi(\XComp) \colon |\Pat| > 1$, which is equivalent to $\Fu{\RACO(V)}(\XComp) \leq \sum_{C \in \Pat} \Fu{\RACO(V)}(C), \forall \Pat \in \Pi(\XComp)$, i.e., $\Fu{\RACO(V)}(\XComp) = \FuHat{\RACO(V)}(\XComp)$. In the same way, one can prove that $\XComp$ is complimentary in the non-asymptotic model if and only if $\Fu{\RNCO(V)}(\XComp) = \FuHat{\RNCO(V)}(\XComp)$. \hfill\IEEEQED

\section{Proof of Lemma~\ref{lemma:SONXComp}}
\label{app:lemma:SONXComp}

The proof is based on \cite[Lemma~\ref{lemma:AlphaAdapt}]{DingITCO2019}. If $\nexists{\XComp} \subsetneq V \colon |\XComp| > 1, \Fu{\alphaU}(\XComp) = \FuHat{\alphaU}(\XComp)$, we must have $\Fu{\alphaU}(\XComp) > \Fu{\alphaU}[\Pat]$ for some $\Pat \in \Pi(\XComp) \colon |\Pat| > 1$ for all $\XComp \subsetneq V$ such that $|\XComp| > 1$. This necessarily means that $\Qat{\alphaU}{V} = \bigwedge \argmin_{\Pat \in \Pi(V)} \Fu{\alphaU}[\Pat] = \Set{\Set{i} \colon i \in V}$. In the case when $\alphaU = \sum_{i \in V} \frac{H(V) - H(\Set{i})}{|V| - 1}$, \cite[Lemma~\ref{lemma:AlphaAdapt}(a)]{DingITCO2019} holds. That is, the PSP of V only contains one critical point $\alphap{1} = \alphaU = \sum_{i \in V} \frac{H(V) - H(\Set{i})}{|V| - 1}$ with the partition chain $\Set{\Set{i} \colon i \in V} = \Patp{1} \prec \Patp{0} = \Set{V}$. In this case, the value of $\alphaU$ in the lemma is in fact the minimum sum-rate for the asymptotic model, i.e., $\alphap{1} = \RACO(V) = \alphaU$, where the necessary and sufficient condition in Corollary~\ref{coro:SOEqv} does not hold, i.e., $\nexists{\XComp} \subsetneq V \colon |\XComp| > 1, \Fu{\RACO(V)}(\XComp) = \FuHat{\RACO(V)}(\XComp)$. Therefore, there is no complimentary subset for SO in the asymptotic model.
Similarly, for $\alphaU' = \big\lceil \sum_{i \in V} \frac{H(V) - H(\Set{i})}{|V| - 1} \big\rceil$ for the non-asymptotic model, we have $\Qat{\alphaU'}{V} = \Set{\Set{i} \colon i \in V} = \Qat{\alphaU}{V}$ due to the property of the PSP $\Qat{\alphaU}{V} \preceq \Qat{\alphaU'}{V}$ for $\alphaU \leq \alphaU'$ (see Section~\ref{subsec:PARReview}). This necessarily means $\alphaU' = \big\lceil \sum_{i \in V} \frac{H(V) - H(\Set{i})}{|V| - 1} \big\rceil = \RNCO(V)$. Corollary~\ref{coro:SOEqv} holds and there is no complimentary subset in the non-asymptotic model. \hfill\IEEEQED

\section{Proof of Lemma~\ref{lemma:SOPAR}}
\label{app:lemma:SOPAR}

    Based on \cite[Lemma~\ref{lemma:EssProp}(b)]{DingITCO2019}, for $\alphaU = \sum_{i \in V} \frac{H(V) - H(\Set{i})}{|V| - 1}$, $\Fu{\alphaU}(C) = \FuHat{\alphaU}(C)$ for all $C \in \Qat{\alphaU}{V_i}$. Also note that we must have $C \subsetneq V_i$ for all $C \in \Qat{\alphaU}{V_i}$. Then, Lemma~\ref{lemma:SOSuf} holds for all $C \in \Qat{\alphaU}{V_i}$ such that $|C| > 1$ in the asymptotic model and therefore there is at lease one complimentary subset. On the contrary, if the partition $\Qat{\alphaU}{V} = \Set{\Set{m} \colon m \in V}$ only contains singleton subsets at the end of last iteration, it means no subset $\XComp \subsetneq V$ such that $|\XComp| > 1$ holds $\Fu{\alphaU}(\XComp) = \FuHat{\alphaU}(\XComp)$. Based on Lemma~\ref{lemma:SONXComp}, there is no complimentary subset in the asymptotic model.
    The same statement for $\alphaU = \big\lceil \sum_{i \in V} \frac{H(V) - H(\Set{i})}{|V| - 1} \big\rceil$ for the non-asymptotic model can be proved in the same way. We prove the optimality of $\rv_{\alphaComp,C}$ as follows.

    In the asymptotic model, consider the value of $\alpha$ satisfying $\Fu{\alpha} (C) = \FuHat{\alpha} (C)$. We have
    $ H(C) + \alpha - H(V) \leq H[\Pat] + |\Pat| (\alpha - H(V)), \forall \Pat \in \Pi(C) $,
    which can be rewritten as $ H(C) -H[\Pat] \leq (|\Pat| -1) (\alpha - H(V))$ and converted to
    $\alpha \geq H(V) - H(C) + \sum_{X \in \Pat}\frac{H(C) - H(X)}{|\Pat| - 1}, \forall \Pat \in \Pi(C) \colon |\Pat| > 1$. Therefore,
    \begin{equation}
        \begin{aligned}
            \alphaComp &= \min \Set{ \alpha \in \Real \colon \Fu{\alpha} (C) = \FuHat{\alpha} (C)} \\
                       &= H(V) - H(C) + \max_{\Pat \in \Pi(C) \colon |\Pat| > 1} \sum_{X \in \Pat}\frac{H(C) - H(X)}{|\Pat| - 1} \\
                       &= H(V) - H(C) + \RACO(C).
        \end{aligned} \nonumber
    \end{equation}
    Also, we have $\rv_{\alphaComp,C} \in B(\FuHat{\alphaComp}^{C},\leq)$ because $\rv_{\alphaComp,C} \in P(\FuHat{\alphaComp}^{C},\leq)$ based on \cite[Lemma~\ref{lemma:EssProp}(a)]{DingITCO2019} and $\rv_{\alphaComp}(C) = \FuHat{\alphaComp}(C) = \FuHat{\alphaComp}^{C}(C)$.\footnote{For an $\alpha$, $\FuHat{\alpha}^{C} \colon 2^{C} \mapsto \Real$ such that $\FuHat{\alpha}^C(X) = \FuHat{\alpha}(X)$ for all $X \subseteq C$ is the \emph{reduction} of $\FuHat{\alpha}$ on $C$ \cite[Section~3.1(a)]{Fujishige2005}.}
    So, the inequality $r_{\alphaComp}(X) \leq \Fu{\alphaComp}(X) = H(X) + \alphaComp - H(V) = H(X) + \RCO(C) - H(C)$ holds for all $X \subseteq C$ and the equality $r_{\alphaComp} (C) = \Fu{\alphaComp} (C) = H(C) + \alphaComp - H(V) = \RACO(C)$ holds for the sum-rate in $C$, i.e., the rate vector $\rv_{\alphaComp,C} \in \RRACO(C)$ is an optimal rate vector that attains the omniscience in $C$ with the minimum sum-rate $\RACO(C)$.
    In the same way, we can prove that $ \alphaComp = \min \Set{ \alpha \in \Z \colon \Fu{\alpha} (C) = \FuHat{\alpha} (C)} = H(V) - H(C) + \RNCO(C)$ and $\rv_{\alphaComp,C} \in \RRNCO(C)$ is an optimal rate vector for the non-asymptotic model. \hfill\IEEEQED

\section{Proof of Corollary~\ref{coro:MstSOACO}}
\label{app:coro:MstSOACO}
    Base on Lemma~\ref{lemma:SOPAR}(a), $\XCompSetk{k}$ obtained in step~\ref{step:MstSOXCompSetk} contains all nonsingleton subsets of $\Patp{p-k}$ that are complimentary. For all $k \in \Set{1,\dotsc,p}$ and each $C \in \XCompSetk{k}$, $\alphaComp = \min \Set{ \alpha \in \Real \colon \Fu{\alpha} (C) = \FuHat{\alpha} (C)} = \alphap{p-k+1}$ so that $\rv_{\alphap{p-k+1},C} \in \RRACO(C)$ with sum-rate $r_{\alphap{p-k+1}}(C) = \RACO(C)$ according to Lemma~\ref{lemma:SOPAR}(a) and Remark~\ref{rem:SOPAR}.

    For $k = 1$, since $\Patp{p} = \Set{\Set{i} \colon i \in V}$, we have $\ASet{C}{\Patp{p}}$ contain only singletons for all $C \in \XCompSetk{1} = \Set{C \in \Patp{p-1} \colon |C| > 1}$. In this case, $\rv_{\alphap{p},V} \in \RRACO(C)$ with the dimensions $r_{\alphap{p},i} \geq 0$ for all $i \in C$ \cite[Lemma~3.23]{Fujishige2005} \cite[Theorem~9]{Ding2018IT}, i.e., the monotonicity in Proposition~\ref{Prop:MstSO}(b) holds. After step~\ref{step:MstSODecomp}, the local omniscience is attained in each $C \in \XCompSetk{1}$ with $r_i^{(1)} = r_{\alphap{p},i}$ for all $i \in C$, i.e., the sum-rate $r^{(1)}(C) = r_{\alphap{p}}(C)$ is assigned to the users in $C$.

    By recursion, before step~\ref{step:MstSODecomp} of Algorithm~\ref{algo:MstSOACO} in iteration $k$, we have all nonsingleton $C' \in \ASet{C}{\Patp{p-k+1}}$ attain local omniscience by an optimal rate vector $\rv_{C'}^{(k')} = \rv_{\alphap{p-k'+1},C'} \in \RRACO(C')$ with sum-rate $r^{(k')}(C') = r_{\alphap{p-k'+1}}(C')$ at some previous stage $k' < k$.
    So, all $C'$ can be treated as super-users with the index $\tilde{C}'$ and, for the problem of attaining the local omniscience in $C$, it suffices to consider the super-user system $\tilde{C} = \Set{\tilde{C}' \colon C' \in \ASet{C}{\Patp{p-k+1}}}$. For $r_{\alphap{p-k+1},\tilde{C}'} = r_{\alphap{p-k+1}}(C') = \sum_{i \in C'} r_{\alphap{p-k+1},i}$, we have $\rv_{\alphap{p-k+1},C} \in \RRACO(C)$ reduce to $\rv_{\alphap{p-k+1},\tilde{C}} \in \RRACO(\tilde{C})$ with $r_{\alphap{p-k+1}}(C) = \RACO(C) = \RACO(\tilde{C}) = r_{\alphap{p-k+1}}(\tilde{C})$.
    Therefore, we just need to assign the rates $r_{\alphap{p-k+1}}(C')$ to the users in $C'$, where the monotonicity in Proposition~\ref{Prop:MstSO}(b) also holds for all $C'$: based on \cite[Lemma~\ref{lemma:EssProp}(c)]{DingITCO2019},
    \begin{equation} \label{eq:DeltaR}
        \begin{aligned}
            \Delta r & = r_{\alphap{p-k+1}}(C') - r^{(k')}(C') \\
                     & = r_{\alphap{p-k+1}}(C') - r_{\alphap{p-k'+1}}(C') > 0
        \end{aligned}
    \end{equation}
    since $\alphap{p-k+1} > \alphap{p-k'+1}$ and $C' \in \Patp{p-k+1}$ so that $\ASet{C'}{\Patp{p-k'+1}} \subseteq \Patp{p-k'+1}$ for all $k' < k$;
    For all singleton $C' \in \ASet{C}{\Patp{p-k+1}}$, we have $r^{(k-1)}(C') = 0$ so that $\Delta r = r_{\alphap{p-k+1}}(C') - r^{(k-1)}(C') \geq 0$ \cite[Lemma~3.23]{Fujishige2005} \cite[Theorem~9]{Ding2018IT}, i.e., Proposition~\ref{Prop:MstSO}(b) holds.
    Note, the above rate updates only need to be considered for all $C \in \XCompSetk{k}$ such that $C \neq \ASet{C}{\Patp{p-k+1}}$. This is because, if $C = \ASet{C}{\Patp{p-k+1}}$, the local omniscience in $C$ has already been attained in the previous stages.

    At the end of the last stage $k = p$, $\rv_{\alphap{p-k+1},C} \in \RRACO(C)$ ensures $\rv_V^{(p)} \in \RACO(V)$ since, in $\XCompSetk{p} = \Set{C \in \Patp{0} \colon |C| > 1} = \Set{V}$, $C = V$ is the only nonsingleton subset and $\rv_{\alphap{1},V} \in \RRACO(V)$.
    Also, because $\Patp{p-k+1} \prec \Patp{p-k}$, $\TXCompSetk{k-1} \subsetneq \TXCompSetk{k}$ for all $k \in \Set{2,\dotsc,p}$. Therefore, $\Set{(\TXCompSetk{k},\rv_{V}^{(k)}) \colon k \in \Set{1,\dotsc,p}}$ is achievable.  \hfill\IEEEQED

\section{Proof of Corollary~\ref{coro:MstSONCO}}
\label{app:coro:MstSONCO}
    It is clear that all $\XCompk{k}$s are complimentary according to Lemma~\ref{lemma:SOSuf} and form the chain $\XCompk{1} \subsetneq \dotsc \subsetneq \XCompk{K} = V$.
    Consider the rate vector $\rvNCO_{\alpha,V}$ returned by the call $\PAR(H,V,\PhiNCO)$. Since $\FuU{\alpha}(\TU{\alpha}{V_i})$ taking integer values for integer-valued entropy function $H$ and $\alpha$ in the non-asymptotic model, we have $\rvNCO_{\alpha,V} \in \Z^{|V|}$ for all integer-valued $\alpha$, i.e., $\rvNCO_{\alphaNCOp{k},V} \in \Z^{|V|}$ for all $k$.
    If $\alphaNCOp{k} = \alphap{j}$, we have shown in the proof of Corollary~\ref{coro:MstSOACO} that $\rvNCO_{\alphaNCOp{k},\XCompk{k}}$ is an optimal rate vector that attains local omniscience in $\XCompk{k}$ with the minimum sum-rate $\RNCO(\XCompk{k}) = \RACO(\XCompk{k})$;
    When $\alphaNCOp{k} \in (\alphap{j},\alphap{j-1}) \cap \Z$, while Lemma~\ref{lemma:SOPAR}(b) states that $\rvNCO_{\alphaNCOp{k},\XCompk{k}} \in \RRNCO(\XCompk{k})$ if $\alphaNCOp{k}  = \min \Set{\alpha \colon \alpha \in [\alphap{j},\alphap{j-1}) \cap \Z}$, it can be proven in the same way that $\rvNCO_{\alphaNCOp{k},\XCompk{k}} \in \RRCO(\XCompk{k}) \cap \Z^{|\XCompk{k}|}$ for any $\alpha \in [\alphap{j},\alphap{j-1})$, i.e., $\rvNCO_{\alphaNCOp{k},\XCompk{k}}$ is achievable, but may not be optimal. Thus, $\rvNCO_{\alphaNCOp{k},\XCompk{k}}$ attains the local omniscience in $\XCompk{k}$.

    For the linear ordering $\PhiNCO$ satisfying \eqref{eq:PhiNCO}, we have $V_{|\XCompk{k}|} = \XCompk{k}$ for all $k \in \Set{1,\dotsc,K}$. Based on \cite[Lemma~\ref{lemma:EssProp}(a)]{DingITCO2019}, the call $\PAR(V,H,\PhiNCO)$ outputs a rate vector $\rvNCO_{\alpha,V}$ such that $\rNCO_{\alpha} (\XCompk{k}) = \rNCO_{\alpha} \big( V_{|\XCompk{k}|} \big) = \FuHat{\alpha}(V_{|\XCompk{k}|}) = \FuHat{\alpha}(\XCompk{k})$ for all $\alpha$. Then, according to \cite[Lemma~\ref{lemma:EssProp}(c)]{DingITCO2019}, for all $k \in \Set{1,\dotsc,K-1}$, since $\alphaNCOp{k} < \alphaNCOp{k+1}$, $\rNCO_{\alphaNCOp{k+1}} (\XCompk{k}) - \rNCO_{\alphaNCOp{k}} (\XCompk{k}) > 0$. \hfill\IEEEQED

\bibliographystyle{IEEEtran}
\bibliography{COSOBIB}

\vfill

\end{document}